\documentclass{article}
\usepackage[utf8]{inputenc}
\usepackage{jheppub}
\usepackage{cleveref}
\usepackage{cancel}
\usepackage{graphicx}
\graphicspath{{figs/}}
\usepackage{ulem}
\usepackage{esint}

\usepackage{floatrow}
\usepackage{placeins} 

\newfloatcommand{capbtabbox}{table}[][\FBwidth]

\newcommand{\Lag}{{\mathcal{L}}}

\newcommand{\Mpc}{{\, {\rm Mpc}}}

\newcommand{\meV}{{\, {\rm meV}}}
\newcommand{\eV}{{\, {\rm eV}}}

\newcommand{\MeV}{{\, {\rm MeV}}}
\newcommand{\GeV}{{\, {\rm GeV}}}

\allowdisplaybreaks

\title{Open String Axiverse
}

\date{}
\author[a]{Rudin Petrossian-Byrne} 
\author[a,b]{and Giovanni Villadoro}
 
\affiliation[a]{Abdus Salam International Centre for Theoretical Physics,
Strada Costiera 11, 34151, Trieste, Italy}

\affiliation[b]{INFN, Sezione di Trieste, Via Valerio 2, I-34127 Trieste, Italy}

\abstract{
Localized charged fields are a general feature of many realistic string compactifications.
In four dimensions they can lead to a multitude of perturbatively-exact global symmetries. If spontaneously broken, they generate a new axiverse compatible with 
post-inflationary evolutions.
}

\begin{document}
\maketitle

%%%%%%%%%%%%%%%%%%%%%%%%%%%%%%%%%%%%%%%%%%%%%%%%%%%%%%%%%%%
\section{Introduction}
\label{sec:Introduction}
%%%%%%%%%%%%%%%%%%%%%%%%%%%%%%%%%%%%%%%%%%%%%%%%%%%%%%%%%%%
%\flag{Note: \cite{xxx} stands for citations to be filled.}

The QCD axion \cite{Weinberg:1977ma,Wilczek:1977pj} is a hypothetical pseudo Nambu-Goldstone boson (NGB) associated to the spontaneous breaking of the 
`Peccei-Quinn' (PQ) $U(1)$ symmetry \cite{Peccei:1977hh,Peccei:1977ur}, anomalous with respect to QCD. It provides arguably the most elegant solution 
to the Strong CP problem \cite{Jackiw:1976pf}. 
Its mass and couplings to the Standard Model (SM) are largely determined by a single parameter, e.g. the axion decay constant $f_a$,
whose current allowed range gives a clear (though challenging) target for detection (see e.g. refs.~\cite{ParticleDataGroup:2024cfk,Adams:2022pbo}).

Expected to contribute to the current energy density of the Universe it could also account naturally for the whole
dark matter abundance observed today \cite{Preskill:1982cy,Abbott:1982af,Dine:1982ah}. The relation between its abundance and its mass strongly depends
on the cosmological history of the axion field, in particular whether after inflation the PQ phase was ever restored. 
In this latter case, a.k.a. `post-inflationary' scenario, the axion decay constant could be predicted.
The calculation is challenging, with the most recent estimates clustering (within an order of magnitude) around $f_a\sim 10^{10}$~GeV \cite{Gorghetto:2020qws,Saikawa:2024bta,Kim:2024wku,Benabou:2024msj} for the minimal 
UV model implementation.

Present bounds on the neutron EDM \cite{Abel:2020pzs} put severe constraints  on the size of extra breakings to $U(1)_{\rm PQ}$ beyond the one from the QCD anomaly \cite{Crewther:1979pi}.
Indeed, in order not to spoil the solution to the strong CP problem, extra breakings at the scale $f_a$ should be smaller than 
$\theta_{\rm QCD}m_u \Lambda_{\rm QCD}^3/f_a^4\sim 10^{-54}$ (for $f_a\sim 10^{10}$~GeV).
A natural question is how such a high-quality symmetry is compatible with the common lore that quantum gravity does not admit exact global symmetries
(a concern sometimes dubbed as the PQ quality problem).
Aside for the short answer ``quantum gravity already breaks PQ via QCD'', a closer inspection to the arguments behind the lore betrays that the true
irreducible breaking of global symmetries are non-perturbative in nature. Hence, weakly coupled UV completions of quantum gravity could allow
for the presence of global symmetries with exponentially high quality at low energies. From this point of view, the PQ quality problem is
more a constraint on the possible UV completions of quantum gravity, rather than an obstruction to low energy physics.

Explicit examples can easily be found in perturbative string theory compactifications \cite{Svrcek:2006yi}, 
which are calculable UV completions of quantum gravity.
There, the zero modes of higher rank gauge fields generically lead to light axion-like particles in four dimensions. The associated axion shift
symmetries (the analogue of our non-linearly realized PQ symmetry) are protected by the higher-dimensional gauge invariance, receiving masses
only from exponentially suppressed contributions by heavy charged objects extending in the compact dimensions. Such constructions 
not only lead to viable QCD axion solutions, but can also easily accommodate the presence of multiple exponentially light axion-like particles (ALPs) besides: a string `axiverse' \cite{Arvanitaki:2009fg}.

It is worth noting that the mechanism behind the axiverse 
can be fully understood within field theory, the only ingredients being
gauge fields and extra dimensions. The simplest prototype  is a $U(1)$ gauge field $A$  compactified on a circle $S^1$ in an extra (5th) dimension \cite{Choi:2003wr}. 
The Wilson loop $\mathcal{W}_{S^1}=\exp{\bigl(ig\int_{S^1}\hspace{-2pt}A\bigr)}$ is identified with a massless
axion field in 4D ($\exp{(i a/v)}\equiv \mathcal{W}_{S^1}$).
The typical axion couplings to topological
charge density operators ($G\tilde G$) in 4D naturally arise from 5D Chern-Simons terms $\varepsilon^{MNRST}A_M G_{NR} G_{ST}$. 
The shift symmetry of the axion\footnote{Descending from the 1-form symmetry $A_M \rightarrow A_M + C_M$ in 5D. See \cite{Craig:2024dnl} for a recent discussion.} is broken in the presence of 5D fields with $U(1)$ charge but, given the non-local nature of the axion in the extra dimension, the breaking is exponentially suppressed ($e^{-ML}$) if the mass $M$ of the lightest charged field is larger than the inverse size $L$ of the extra dimension. While in field theory this is only a plausible accident, in string theory constructions it is a generic feature. 

% explicit string theory constructions justify the assumption.

From the phenomenological point of view, a potential shortcoming of 
the string axiverse lies in the challenges to implement
the more predictive post-inflationary scenario. The latter would require a mechanism to produce 
axion strings, which in the string axiverse would 
correspond to (possibly a bound state of) D and/or NS branes.
The efficient production of such objects poses a theoretical, if not phenomenological, challenge,
since an explicit computation would require entering non-perturbative string regimes in a cosmological setup (see, e.g., ref.~\cite{March-Russell:2021zfq,Benabou:2023npn,Reece:2024wrn} for  recent discussions about the topic).

\begin{figure}[t]
    \centering
    \includegraphics[scale=0.3]{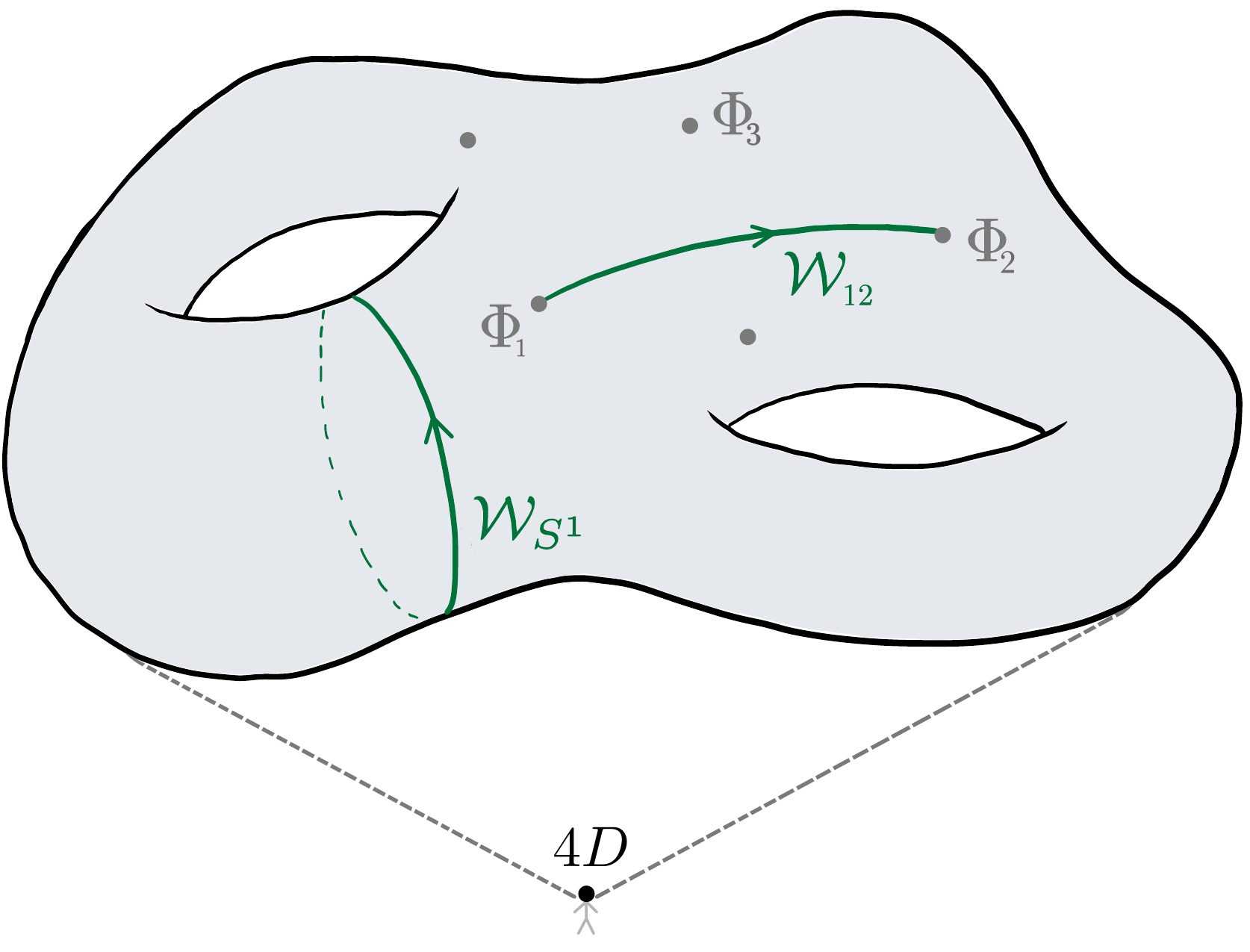}
    \caption{An exponentially good global symmetry in 4D can be naturally obtained when the only gauge-invariant objects (green in figure) charged under it are non-local in the extra dimension(s). When spontaneously broken, an exponentially light axion appears in the IR. 
   To the left, the older case of a Wilson loop $\mathcal{W}_{S^1}$ wrapping a circle, charged under the shift symmetry of a gauge field.
   To the right, the line operator $\mathcal{W}_{12}$ connecting two spatially separated fields $\Phi_{1,2}$, charged under simultaneous rotations orthogonal to the gauged combination. 
   % Grey dots denote the position of matter fields. In type I string theory constructions,  these are intersection points between D5 branes. 
   } 
    \label{fig:sketch}
\end{figure}

In this work, we discuss a different way to realize light axions in higher-dimensional/string-theory compactifications.
The mechanism just requires the presence of multiple charged fields localized in different positions of the compact manifold; the axion
then emerges from extended gauge-invariant non-local operators in the extra dimension, as sketched in \cref{fig:sketch}. Unlike the previous case, the axions
 are mostly associated to the open string sector of the theory, and therefore decoupled from the gravitational sector and the string scale.
This mechanism is consistent with a full four dimensional restoration
of the PQ phase, allowing the post-inflationary QCD axion to enjoy the same amount of PQ protection as its closed string cousins.
In fact, we will show that this mechanism could also be generic in a large class of string compactifications, suggesting the presence
of multiple light ALPs beyond the QCD axion, possibly also experiencing post-inflationary evolution --- an {\it open string axiverse}.
While sharing similar properties, the open string axiverse may present some phenomenologically new opportunities, given the calculability
of the axion abundance, potential gravitational signals deriving from the topological defects produced from the PQ phase transition, as well as
dark matter substructures such as mini-halos and Bose stars. 

As for the string axiverse, the mechanism behind the open string
axiverse can be fully understood within field theory. It can naturally 
be extended to protect any $U(1)$ global symmetry with exponentially high quality and it requires adding only a small extra dimension. 
An important point is that the high-quality global symmetry does not 
manifest itself in the effective 4D theory as an accidental symmetry,
rather as a genuine global symmetry: symmetry breaking operators of low
dimensionality are not forbidden, but their coefficient is exponentially suppressed.
This reconciles the ancient QFT practice of imposing global symmetries
by hand in low-energy effective field theories (EFTs) with the modern lore about 
the absence of exact global symmetries in a full theory of quantum gravity.

It is worth mentioning that some of the ideas discussed in the present work have already been used in the past. In particular the importance of sequestering to generate approximate symmetries in extra dimensions was realized long ago in ref.~\cite{Arkani-Hamed:1998lzu} and applied in a specific model of QCD axion to protect the PQ symmetry in ref.~\cite{Cheng:2001ys}. A similar idea in string theory for generating high-quality global $U(1)$ symmetries was introduced in ref.~\cite{Ibanez:1999it} and applied to generate open string axions in various 
contexts (see e.g. ~ref.~\cite{Berenstein:2012eg,Honecker:2013mya,Cicoli:2013cha,Choi:2014uaa} and references therein). 
% Further comparison between this work and the literature is deferred to \cref{sec:Conclusions}.

The structure of the paper is as follows. In \cref{sec:AxionsFromWilsonLines}
we describe the basic mechanism at the level of higher-dimensional gauge field theories, in \cref{sec:5DOrbifold} we discuss an explicit 5D construction
that could reproduce the minimal KSVZ QCD axion model with high quality and discuss
various extensions in \cref{sec:modelbuilding}. In \cref{sec:StringTheory}
we discuss how to embed the mechanism in string theory and how the open string
axiverse emerges. In \cref{sec:cosmology} we overview potential cosmological and phenomenological implications. Our conclusions are given in
\cref{sec:Conclusions}. Finally we provide some
additional details about the 5D constructions in \cref{app:5DOrbifold}, more complete examples of string theory embeddings --- discussing
the constraints from supersymmetry and consistency conditions --- in \cref{app:susystring}, and details of exotic cosmic string solutions in \cref{app:string_solutions}.

%%%%%%%%%%%%%%%%%%%%%%%%%%%%%%%%%%%%%%%%%%%%%%%%%%%%%%%%%%%
\section{Symmetries and Axions from Separated Charges}
% Field theory axion with extra-dimensional protection.
\label{sec:AxionsFromWilsonLines}
%%%%%%%%%%%%%%%%%%%%%%%%%%%%%%%%%%%%%%%%%%%%%%%%%%%%%%%%%%%
We start by discussing how gauge theories in extra dimensions allow for high-quality global symmetries and axions  in the presence of sequestered charged fields.
% \footnote{The same mechanism has already been applied in a specific 5D model to realize a well protected
% QCD axion in \cite{Cheng:2001ys}, which has partial overlap with the discussion in this section.}.
We illustrate this by considering the simplest case of a $U(1)$ gauge theory and two charged fields $\Phi_\pm$ spatially separated at coordinate points $y_\pm$ (with $|y_+-y_-|=L$) in the extra dimension(s). 
The precise nature of these is not crucial; we look at  a concrete 5D orbifold construction in the following subsections, and string theory in \cref{sec:StringTheory}.
For simplicity, we take the charges with respect to the bulk $U(1)$ to be equal and opposite, $\pm q $ respectively.
% \footnote{Matching onto a particular string theory picture  in \cref{sec:StringTheory} will give $q=2$, but this does not change the physics.} 
 Under gauge transformations,
\begin{align}
    A_M \rightarrow A_M +\partial_M \Lambda \ , \quad \Phi_\pm \rightarrow \Phi_\pm e^{\pm i g_d q\Lambda_{\pm}} \ , 
\end{align}
% \begin{align}
%     A_M \rightarrow A_M -\partial_M \Lambda \ , \quad \Phi_- \rightarrow \Phi_- e^{-i g_5 \Lambda(0)} \ , \quad \Phi_+ \rightarrow \Phi_+ e^{i g_5 \Lambda(L)} \ ,
% \end{align}
where $M=\mu,5\dots$ runs over spacetime dimensions, $g_d$ is the extra dimensional gauge coupling, and  $\Lambda_\pm(x^\mu) \equiv \Lambda|_{y=y_\pm}$. The two matter fields transform completely independently. 
From the limited perspective of each brane, both $U(1)_{\pm}$ symmetries of phase rotations of $\Phi_\pm$ are gauged, and thus cannot be explicitly broken. On the other hand, only one linear combination of rotations corresponds to the global ($\Lambda = \rm const$) symmetry that is gauged. The orthogonal transformation of phase rotations $\Phi_{\pm} \rightarrow \Phi_\pm e^{i \alpha}$, $\alpha \in \mathbb{R}$, is an independent symmetry that we will tellingly refer to as PQ. Like any symmetry, this can (and we expect it to) be broken explicitly. However, similar to the more familiar string axiverse case, when descending to 4D, it will be afforded a certain level of protection due to gauge redundancy in the extra dimension, because 
the only gauge-invariant objects charged under it are necessarily non-local, such as the line operator
% \flag{XXX}
\begin{align}
\label{eq:WilsonLine}
    \mathcal{W} = %\Phi_- \, \exp{\left(-i g_5 q \int_{0}^{L} \, dy \, A^y\right)} \Phi_+  
    \Phi_+ \, e^{i g_d q \int_{y_+}^{y_-} dy \, A_y }\, \Phi_-    \ .
\end{align}
% \sout{where we took a flat metric for simplicity.}
Breaking the PQ symmetry, and thereby generating non-local terms like \cref{eq:WilsonLine} in the effective action, requires some charged bulk field coupled to both brane-localized fields.
As an example in 5D, we can try breaking PQ by the most relevant possible interaction  $\Lag \supset -M^2 |\varphi|^2+  \mu_-^{3/2}\varphi\, \Phi_- \delta(y-y_-) + \mu_+^{3/2} \varphi^\dagger\,\Phi_+ \delta(y-y_+) + {\rm h.c.}$,
with $\varphi$ some heavy charge-$q$ bulk scalar field connecting $\Phi_\pm$, with mass $M L \gg 1 $, and $\mu_{\pm}$ some dimensionful parameters. It is easy to show that integrating out $\varphi$ produces the leading 4D potential
\begin{align}
\label{eq:PQbreaking}
   V_{\cancel{\rm PQ}} = \frac{(\mu_-\mu_+)^{\frac{3}{2}}}{ 2 M}\,e^{-ML}\, {\cal W}+ {\rm h.c.} + \dots
\end{align}
Thus, as usual, as long as charged bulk states are somewhat heavier than the KK scale, we have an exponentially good global symmetry. Note that,
from a 4D perspective, no deep principle forbids the relevant PQ-breaking operator $\Phi_-\Phi_+$ in the EFT. It is not an accidental symmetry. Instead, non-locality in the extra dimension(s) explains the exponentially suppressed coefficient. In 4D, PQ appears as a genuine global symmetry imposed by hand.

% Compare with other models.
% In the 4d theory there is nothing forbidding $\Phi_-\Phi_+$, which would break PQ symmetry. But

We can now imagine spontaneously breaking $U(1)_\pm$ at $y=y_\pm$. If $\Phi_{\pm}$ are indeed scalars, this is easily achieved through negative mass terms in their brane-localized potentials $V_\pm(|\Phi_{\pm}|)$, leading to vacuum expectation values $\langle |\Phi_{\pm}|\rangle=v_\pm/\sqrt{2}$. 
Without the gauge field, this would lead to two NGBs. Instead, one linear combination is eaten by the gauge field, while the other, gauge-invariant combination, remains as an axion. Without the extra dimension, the latter is just $\arg\left(\Phi_-\Phi_+\right)$. 
But in 5D this is not gauge invariant.
The axion is properly identified with the phase of the non-local gauge-invariant line operator
\begin{align}
\label{eq:AxionDef}
    a(x^\mu)/v \equiv \arg (\mathcal{W}) \ ,
\end{align}
and is thus a linear combination of (the phases of) the $\Phi_\pm$ and the extra component of the gauge field $A_y$. Throughout this work, we define $v$ so the fundamental domain of an axion is $2\pi v$. 

PQ breaking, such as \cref{eq:PQbreaking}, would contribute to the potential for this axion, which can thus be exceedingly light.
To make it the QCD axion we just need the global symmetry to be anomalous under the $SU(3)_c$ color group of strong interactions. This is easily done by the inclusion of chiral colored fermions, as  shown explicitly in a concrete 5D model below, and in string theory in \cref{sec:StringTheory}. To properly solve the Strong CP problem, the breaking from the QCD anomaly at low energy should by far dominate over those from eq.~(\ref{eq:PQbreaking}).

It is of course straightforward to generalize to an arbitrary number $N$ of spatially separate$D$ branes at $y=y_i$, and $N$ fields $\Phi_i$ localized thereon with generic charges $q_i$, giving rise to
$N-1$  global symmetries. These are simultaneous constant phase rotations of the $\Phi_i$ with charge vectors orthogonal to the gauged $q_i$. 
Again, the only gauge-invariant objects charged under the global symmetries are non-local, such as the lines operators connecting the various fields, $\mathcal{W}_{ij} = \Phi_i^{p_{i}} \, e^{\left(i g_5 m_{ij} \int_{y_i}^{y_j} \, dy \, A_y\right)} (\Phi^{\dagger}_j)^{p_{j}} $,
where $m_{ij}$ is the lowest common multiple of  $q_{i},q_{j}$ charges, $m_{ij}=p_iq_i=p_jq_j$, and if $p_i$ is negative one should interpret $\Phi^{p_{i}}$ as $(\Phi^\dagger)^{|p_{i}|}$. Of course, only $N-1$ of the ${\cal W}_{ij}$ are independent.

\begin{figure}[t]
        \centering
         \includegraphics[width=0.7\textwidth]{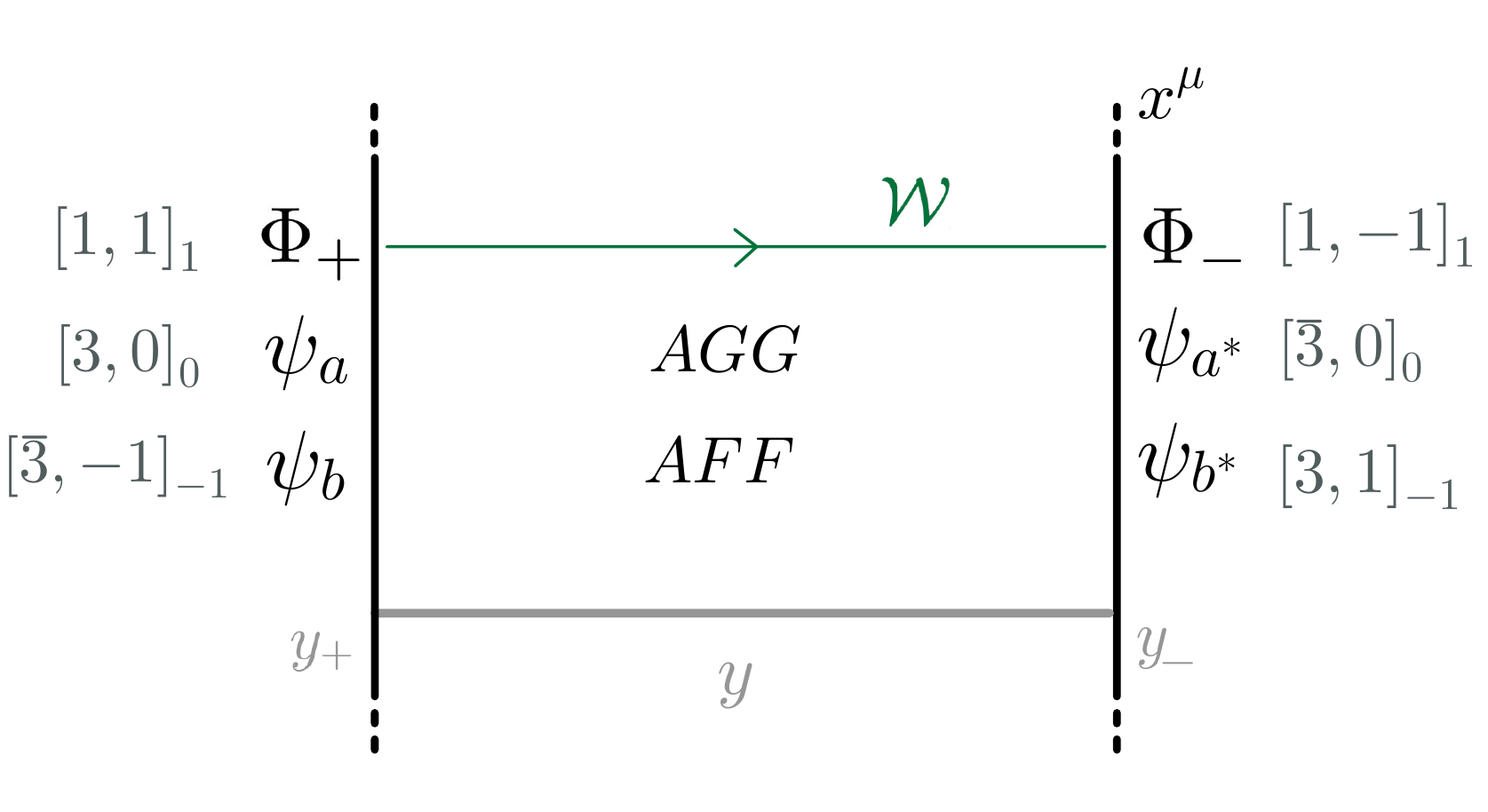}
         %
         %%%%%%%%%%% Caption %%%%%%%%%%
        \caption{Schematic of a 5D orbifold model for the QCD axion, with table of matter content and representations under $\left[SU(3)_c,U(1)_5\right]_{U(1)_{\rm PQ}}$. Matter fields live on the $y=y_\pm$ branes as depicted. The axion is the phase of the (green) line operator $\mathcal{W}$ connecting $\Phi_\pm$. The sketched Chern-Simons terms are required in the bulk to communicate anomaly cancellation. In the \textit{KK}SVZ limit, fields on one brane can be replaced by different boundary conditions.}
        %%%%%%%%%%%%%%%%%%%%%%%%%%%%%%%
        %
        \label{fig:5DorbifoldSketch}
\end{figure}
%%%%%%%%%%%%%%%%%%%%%%%%%%%%%%%

% \begin{align}
%     & \left[1, &-1 \right]_1 \\
%     & \left[3, &0 \right]_{0}  \\
%      & [\bar{3}, &-1 ]_{-1}
% \end{align}

%%%%%%%%%%%%%%%%%%%%%%%%%%%%%%
\subsection{5D for the axion of QCD}
\label{sec:5DOrbifold}
%%%%%%%%%%%%%%%%%%%%%%%%%%%%%%

To be a bit more concrete, we can take the position $y_\pm$ of the 4D localized charged fields $\Phi_\pm$ to be boundary branes of an interval comprising a single (5th) flat extra dimension\footnote{An $S^1/\mathbb{Z}_2$ orbifold construction, for the extra-dimensional aficionados.}.
 The Abelian gauge group above, now dubbed $U(1)_5$, and QCD $SU(3)_c$ live in the bulk, with Neumann (Dirichlet) boundary conditions imposed on their $\mu$ (5th) components. 
We now take $q= 1$ for simplicity.
With $\langle \Phi_\pm \rangle = v_\pm /\sqrt2$, upon dimensional reduction to 4D, 
$A_M$ produces a tower of massive vector fields whose transverse polarizations are provided by the KK modes of the 4D components $a_\mu^n$ while the longitudinal by linear combinations of the KK modes of the 5th component $a_5^n$ and the two phases of the localized fields $\theta_\pm = \arg(\Phi_\pm) $. The following gauge-invariant linear combination of fields remain as a NGB,
%$A^M$ becomes a tower of KK modes: $N_K$ vectors $a^\mu_n$ and $N_K-1$ NGBs $a^5_n$, plus the two $\theta_\pm = \arg(\Phi_\pm) $ arising from SSB at the boundaries. An appropriate gauge transformation $A_M \rightarrow A_M - \partial_M \Lambda$ can be used to diagonalize the tower of $N_K+1$ NGBs leaving only the gauge invariant combination
\begin{align}
\label{eq:AxionOrbifold}
    a/v = \theta_- + & \theta_+ - 2 \sqrt{2} \, g_4\sum_{n \;{\rm odd} }{L \over n\pi}a_5^{n} \ ,  \qquad 
    v^2 = \left( { g_4^2 L^2}
    + {1\over v_-^2} + {1\over v_+^2} \right)^{-1} \ , 
\end{align}
where $g_5=g_4\sqrt{L}$ is the relation between 4D and 5D gauge coupling.
The combination in \cref{eq:AxionOrbifold} is more simply read off from the manifestly gauge-invariant definition in \cref{eq:AxionDef} as the argument of the line operator (\ref{eq:WilsonLine}). 
%%%%%%%%%%%%%%%%%%%%%%%%%%%%%%%%%%%%%%%%%%%%%%%%%%%%%%%%%%%%%%%
%%%%%%%%%%%%%%%%%%%%%%%%%%%%%%%%%%%%%%%%%%%%%%%%%%%%%%%%%%%%%%%
% The sum is only over positive odd $n$ because \flag{xxx}.
%%%%%%%%%%%%%%%%%%%%%%%%%%%%%%%%%%%%%%%%%%%%%%%%%%%%%%%%%%%%%%%
%%%%%%%%%%%%%%%%%%%%%%%%%%%%%%%%%%%%%%%%%%%%%%%%%%%%%%%%%%%%%%%
Notice that $v$ is controlled by the lightest scale in the problem. More details are provided in \cref{app:5DOrbifold}.

To make \cref{eq:AxionOrbifold} the QCD axion we add quarks to both branes as depicted in \cref{fig:5DorbifoldSketch}, where the transformation properties of all fields are tabulated as well. This allows for brane-localized Yukawa couplings 
${\rm y}_+\Phi_+ \psi_{a} \psi_{b}  + {\rm h.c.}$ and  ${\rm y}_-\Phi_- \psi_{a^\star} \psi_{b^\star} + {\rm h.c}$.
The theory on each brane looks like the KSVZ model \cite{Kim_KSVZ_1979,SVZ_of_KSVZ_1980}, arguably the operationally simplest 4D UV completion for the QCD axion, except for the extra $U(1)_5$ gauging. Although the theory  as a whole is vector-like and (gauge) anomaly free in 4D, two Chern-Simons terms are required in the bulk to communicate the canceling of the cubic $U(1)_5^3$ and mixed $U(1)_5SU(3)_c^2$ brane-localized gauge anomalies, 
\begin{align}
\label{eq:ChernSimons}
        S_{\rm UV} \supset \int d^5x 
   \, \left\{ { 3  g_5^3 \over 32 \pi^2} \, \epsilon^{MNRST}  A_M F_{NR}F_{ST} + { g_5 g_{s,5}^2 \over 64 \pi^2} \, \epsilon^{MNRST} A_M G^a_{NR}G^a_{ST}
    \right\} \ ,
\end{align}
where $G$ is the field strength of QCD and $g_{s,5}$ is the 5D strong coupling.
The global PQ symmetry is instead anomalous with respect to both gauge groups.
The $U(1)_{\rm PQ}SU(3)_c^2$ anomaly leaves its imprint in the 4D IR by the defining interaction of the QCD axion
\begin{align}
\label{eq:QCDaxion_defining_interaction}
    S_{\rm IR} \supset \int d^4x \, {g^2_{s} \over 32 \pi^2}\left({a\over f_a} - \theta_{\rm QCD}\right) G^a_{\mu\nu}{\tilde{G}^{a,\mu\nu}} \ , \qquad  f_a \equiv  v/N_w \ ,
\end{align}
where  $\tilde{G}$ is the dual field strength, and $\theta_{\rm QCD}$ absorbs any remnant of CP violation.
In our KSVZ-like model of \cref{fig:5DorbifoldSketch}, the coefficient $N_w$ counts the number of PQ-charged colored fermion pairs on each brane, with $N_w=1$ the minimal case highlighted.\footnote{This can be seen most straightforwardly by the fact that, under a PQ transformation $\Phi_\pm \rightarrow \Phi_\pm e^{i \alpha }$, etc., both $a/v$ (by the definition in \cref{eq:AxionOrbifold}) and $\theta_{\rm QCD}$ term (by the anomaly) shift by the same amount $2\alpha$. }
More generally, it denotes the discrete $\mathbb{Z}_{N_w}$ symmetry preserved by the anomalous breaking of $U(1)_{\rm PQ}$ by QCD.
In cosmology (see \cref{sec:cosmology}), this is directly related to the minimum number of domain walls attached to axion strings produced in the post-inflationary scenario, with potentially dangerous consequences for $N_w>1$. 

From the perspective of the 5D EFT, extra sources of PQ breaking can arise by the effects of charged bulk states as in the example of \cref{eq:PQbreaking} above. These can easily be made consistent with the non-trivial experimental constraint $V_{\cancel{\rm PQ}}\lesssim 10^{-10}m_u \Lambda_{\rm QCD}^3$ thanks to the exponential suppression $e^{-ML}$, which can be as large as $e^{-24\pi^3/g_s^2}\sim 10^{-{320}}$ for $g_s=1$, taking $M$ at the cut-off of the 5D theory $\Lambda_5 \approx 24\pi^3/L g_s^2$ by NDA \cite{Chacko:1999hg}. This level of protection is the same as in the more familiar toy example of a Wilson loop axion $a/v = \arg(\mathcal{W}_{S^1})$ discussed in the introduction. 
In any case, in field theory the presence of charged states with $M<\Lambda_5$ is not compulsory, and one can imagine the true irreducible breaking effects from quantum gravity being even smaller.
A sharper quantitative measure of minimal breaking/protection is only calculable in a UV completion such as string theory, as explored in \cref{sec:StringTheory}.

% \begin{figure}
% \begin{floatrow}
% \ffigbox{%
%   \includegraphics[scale=0.13]{figs/5DorbifoldSketch2.png}%
% }{%
%   \caption{A figure hasoifhatfoaufpoaugf9aumfoapmf jisa saija afjoiaf jajfajsaf jao'pow oiu ar'aspof ;l'  'lj fsa;o'asfj'asf;o'af lkaslk'fslkjfsaiusa safjoi'sa fsoijhf lsamfmxosuf oijslkfjsalf jsaoijhf;oisa hfsanflkndlk fj oidjf lsf  fsjf;sjf}%
% }
% \capbtabbox{%
%   \begin{tabular}{||c || c c c | c c c||} 
%  \hline
% & $\Phi_-$ & $\psi_L$ & $\chi_R$ & $\Phi_+$ & $\psi_R$ & $\chi_L$ \\ [0.5ex] 
%  \hline\hline
% $U(1)_5$ & -1 & -1 & 0 & 1 & -1 & 0 \\ 
%  \hline
% $SU(3)_c$ & 1 & 3 & 3 & 1 & 3 & 3 \\
% \hline
% $U(1)_{\rm PQ}$ & 1 & 1 & 0 & 1 & -1 & 0 \\
%  \hline \hline
% $U(1)_Y$ & 0 & 2/3 & 2/3 & .  & . &  . \\
% \hline
% \end{tabular}
% }{%
%   \caption{A figure hasoifhatfoaufpoaugf9aumfoapmf jisa saija afjoiaf jajfajsaf jao'pow oiu ar'aspof ;l'  'lj fsa;o'asfj'asf;o'af lkaslk'fslkjfsaiusa safjoi'sa fsoijhf lsamfmxosuf oijslkfjsalf jsaoijhf;oisa hfsanflkndlk fj oidjf lsf  fsjf;sjf}%
% }
% \end{floatrow}
% \end{figure}

% \vspace{0.5cm}
% \flag{Comments:
% \begin{itemize}
%     \item One option would be to put this explicit model in appendix.
% \end{itemize}
% }

\subsection{Variations on a theme}
\label{sec:modelbuilding}
The particular setup highlighted here should be viewed as a benchmark model, one that admits exponentially good PQ quality and the post-inflationary scenario with $N_w=1$. Of course, as with the usual KSVZ model in 4D, the field content can be extended or replaced in many ways to address problems, perceived weaknesses, or unify the axion with other BSM physics (see e.g. ref.~\cite{DiLuzio:2020wdo}). These can be pursued without interfering with the protection mechanism.

%%%%%%%%%%%%%%%%%%%%%%%%%%%%%%
\subsection*{Scale separation}
\label{sec:Naturalness}
%%%%%%%%%%%%%%%%%%%%%%%%%%%%%%

A valid concern is the use of fundamental scalars, with their associated tuning. For us, this is evidently true when separating the axion decay constant from the KK scale $v \ll L^{-1}$, where the protection mechanism takes place.
This hierarchy problem can be addressed in the usual ways, either by making the axion a composite state of confining dynamics, or by supersymmetry (SUSY). Composite axion models \cite{Kim:1984pt,Choi:1985cb} tend to have $N_w >1$, which could be more problematic in a post-inflationary scenario for the QCD axion, and we therefore focus here on the SUSY extensions.
Consider the 4D boundary theories to have $\mathcal{N}=1$ SUSY, promoting to chiral superfields $\Phi_\pm \rightarrow \hat{\Phi}_\pm$ etc. 
% \sout{(in 5D, where the gauge fields live, the minimal number of supercharges is double, but this will not enter our discussion)}. 
To justify the desired hierarchy, the scale of SUSY breaking can be at or (parametrically) below $v$. We highlight two simple scenarios. 

\paragraph{SUSY I:} To spontaneously break symmetries while preserving SUSY requires more fields: on each brane, an additional chiral superfield $\hat{\Phi}'_\pm$ with opposite $U(1)_5$ charge to $\hat{\Phi}_\pm$, and a neutral $\hat{S}_\pm$. The Mexican hat is replaced by the superpotential
 $W|_{y=y_\pm} \supset \lambda_\pm \hat{S}_\pm (\hat{\Phi}_{\pm} \hat{\Phi}'_\pm - {v_\pm^2 / 2} ) + w_\pm(\hat S_\pm)$,
% \begin{align}
%     \left. W\right|_{y=y_\pm} \supset \lambda_\pm \hat{S}_\pm \left(\hat{\Phi}_{\pm} \hat{\Phi}'_\pm - {v_\pm^2 \over 2} \right) + w_\pm(\hat S_\pm)\ ,
% \end{align}
where $w_\pm(\hat{S}_\pm)$ are some (cubic) polynomials. For mild assumptions on $w_\pm$, the scalar potential is minimized by the SUSY-preserving, $U(1)_5$-breaking vacuum $\langle \Phi_\pm \Phi'_\pm \rangle = v_\pm^2/2$ and $\langle S_\pm\rangle = 0$, where here the lack of hats denotes the scalar part\footnote{The extra flat direction does not affect the physics of interest here and is ultimately lifted by SUSY-breaking effects. Adding even more charged fields, one can also imagine simultaneously breaking SUSY and $U(1)_5$ simultaneously \cite{HARIGAYA2017507}. }. 
While the combination of phases ${\arg}\left(\Phi_{\pm} \Phi'_{\pm}\right)$ get a mass of order $\lambda_\pm v_\pm$, the orthogonal is identified with $\theta_\pm \equiv {\arg}\left(\Phi_\pm \Phi'^\dagger_\pm\right)$ in \cref{eq:AxionOrbifold}.

\paragraph{SUSY II:} Without the introduction of additional fields, the scalar potential not involving colored scalars comes solely from the $U(1)$ D term and is proportional to $g_4^2 \left( |\Phi_+|^2-|\Phi_-|^2\right)^2$. We can imagine PQ spontaneously broken at the same scale that SUSY-breaking soft terms lift the flat direction, giving vacuum expectation values (vevs) $\langle|\Phi_+|\rangle \approx \langle|\Phi_-|\rangle \simeq  v \gg m_{\rm soft}$, with $m_{\rm soft}\sim F/M_{\rm mess}$ the soft SUSY-breaking tachyonic mass term, $F$ being the SUSY breaking order parameter and $M_{\rm mess}$ the messenger scale.

\paragraph{} In both cases, the axion gains the label of QCD by introducing chiral superfield versions of the colored fermions, with superpotential terms $ W|_{y=y_\pm} \supset  {\rm y}_{\pm}\hat{\Phi}_\pm \hat{\psi}_{a^{(\star)}} \hat{\psi}_{b^{(\star)}} $.
While these introduce new scalars into the potential, they do not complicate the vacuum properties stated above and their vevs are zero.

\subsection*{Relic decays}
In the case of the post-inflationary scenario, an extension is phenomenologically necessary to avoid thermal populations of the heavy colored fermions $\psi$ overclosing the universe / giving rise to stable  (electrically) fractionally charged hadrons. Extensions that preserve $N_w=1$ simply give $\psi_{a^{(\star)}}$ the same quantum numbers as the Standard Model $u^c$ or $d^c$ (hypercharge $Y=-2/3$ or $1/3$ respectively) allowing them to decay to the SM thanks to the ensuing mixing \cite{DiLuzio:2016sbl}. For example, the Yukawa interaction can be extended to $\Phi_- \left( {\rm y_-} \psi_{a^\star}+{\rm y}'_{-}d^c\right)\psi_{b^\star} + {\rm h.c.}$, and once $\langle \Phi_{-} \rangle \neq 0 $, the heavy mass eigenstate $({\rm y}_- \psi_{a^\star}+{\rm y}'_{-}d^c)/\sqrt{{\rm y}_-^2+{\rm y}_{-}'^2}$ can decay to a light quark and Higgs through SM Yukawa interactions. 
% The mixing with the SM fermions can easily be small enough not to have observable effects yet large enough to make the charged heavy fermions sufficiently unstable. 
Assigning a non-zero hypercharge to the localized fields
creates local gauge anomalies that need to be canceled by 5D bulk Chern-Simons terms analogous to those in eq.~(\ref{eq:ChernSimons}). One phenomenologically relevant implication is that the coupling of the axion to photons will differs 
from the one used for the minimal KSVZ benchmark model. 
Note however that this type of extensions is intrinsically four dimensional (as we are mostly adding light degrees of freedom on the branes), therefore it shares the same properties and phenomenology of the one already studied in the literature \cite{DiLuzio:2016sbl}.

\subsection*{Grand Unification}
The intrinsic extra-dimensional nature of the PQ symmetry does not interfere 
with theories of Grand Unification (GUT). 
This is because the exponential protection of the PQ symmetry
requires only a small extra dimension that can lie above the expected
GUT scale ($1/L\gtrsim v_{\rm GUT}\sim {\cal O}(10^{16})$~GeV). The GUT gauge fields (e.g. $SU(5)$) would
live in the bulk of the extra dimension together with the extra $U(1)$, 
while the fields localized on the brane would now involve complete representations of the GUT group. In particular for the specific model presented before $\psi_{a^{(\star)},b^{(\star)}}$ could be uplifted to (anti-)fundamental of $SU(5)$ while $\Phi_{\pm}$ remain singlets.

%A reasonable theory bias for the UV is grand unification (GUT). The model in \cref{sec:5DOrbifold} can minimally be made consistent with GUT by upgrading $SU(3)_c$ in the bulk to the GUT group (e.g. $SU(5)$), $\psi_L,\chi_R$ to fundamentals thereof, and taking $L^{-1}$ above the unification scale, while the gauge group $U(1)_5$ is kept as an extra factor.
% \flag{mention Prateek, Mario paper?}

%%%%%%%%%%%%%%%%%%%%%%%%%%%%%%
\subsection*{\textit{KK}SVZ}
\label{sec:KKSVZ}
%%%%%%%%%%%%%%%%%%%%%%%%%%%%%%

In the regime $v\simeq v_-\ll v_+, L^{-1}$, at low energies one does not see the degrees of freedom on the $+$ brane, nor the $U(1)_5$ gauging, and the previous setup reduces to the familiar benchmark 4D model of KSVZ  \cite{Kim_KSVZ_1979,SVZ_of_KSVZ_1980}. While the phenomenology becomes indistinguishable, the protection remains due to the axion being in reality an extended object in the 5th dimension, partly comprised of KK modes of the gauge field. Thus, we deem `\textit{KK}SVZ' an appropriate name. In this manner, the PQ quality  can be justified even for the simplest UV axion model, where the global symmetry is simply declared by hand.

It is straightforward now to realize a post-inflationary scenario, if the maximum temperature of the universe is high enough ($T_{\rm max} > v$) to restore
PQ but not high enough to excite heavier degrees of freedom ($T_{\rm max} < v_+,L^{-1}$). The axion strings produced look just like regular field theory solutions and the string network evolution will follow the same dynamics. More complicated scenarios are of course possible, but we delay their study to \cref{sec:cosmology}. 
% \flag{Mention here something about string in SUSY version of KKSVZ or leave for Cosmo sec?}

\subsection{KK-lifting of global symmetries}
We note that the formal limit $v_+\rightarrow\infty$, completely decoupling the fields on the right brane, can be obtained by a shortcut, imposing Dirichlet (Neumann) boundary conditions on $A_\mu$ ($A_5$) there, instead of the original, inverse ones\footnote{Amounting to an $S^1/\mathbb{Z}_2\times \mathbb{Z}_2$ orbifold construction, for the extra-dimensional enthusiasts. Notice that the decoupling limit $v_\pm \rightarrow \infty$ is akin to flipping the boundary conditions on \textit{both} branes, and we recover the older case of the extra-dimensional axion identified with the zero mode of $A_5$. }.
In this way the KKSVZ axion-defining non-local line operator becomes simply $ a(x^\mu)/v\equiv {\rm arg}\bigl(\,e^{i g_5\!\int_{y_+}^{y_-}\! dx^5\! A_5}\Phi_-\bigr)  $.
This procedure, one might call `KK-lifting', can be performed on any conceivable 4D UV axion model, and in fact more generally to protect any $U(1)$ global symmetry 
up to non-perturbatively small effects.  The recipe reads:
\begin{enumerate}
    \item Gauge the global 4D (PQ) $U(1)$ symmetry;
    \item Extend the new gauge field into the bulk of a finite length 5th dimension, including an appropriate Chern-Simons term therein if the original symmetry was anomalous;
    \item Impose Dirichlet (Neumann) boundary conditions on its $\mu$ (5th) components on the other side of the extra dimension.
\end{enumerate}
Any 4D local operator ${\cal O}_q(x^\mu)$ with charge $q\neq 0$ under the global $U(1)$ symmetry would be non gauge invariant from the 5D point of view unless it gets
dressed by the appropriate Wilson line, i.e. $ e^{iqg_5\!\int_{y_+}^{y_-}\! dx^5\! A_5} {\cal O}_q(x^{\mu})$. Such non-local operators could only be generated from charged fields living
in the extra dimension with coefficients that are exponentially suppressed as long
as they are heavy enough.

The further generalization to non-Abelian symmetries is left as an exercise to the reader.

 % can be uplifted to its protecting extra-dimensional setup by imagining that the action of PQ transformations on the 4$D$ brane theory is actually gauged. The corresponding gauge field (as well as QCD) however extend into the bulk of the extra dimension. In the 5D picture PQ transformations are not gauged, and so the QCD axion survives uneaten.

%%%%%%%%%%%%%%%%%%%%%%%%%%%%%%%%%%%%%%%%%%%%%%%%%%%%%%%%%%%
\section{Symmetries and Axions from Open Strings}
\label{sec:StringTheory}
We now illustrate how the basic ingredients required to produce
high-quality (spontaneously broken) symmetries at low energy are not
only possible but even generic in a large class of string theory compactifications. 
Indeed, besides extra dimensions and gauge theories, localized charged
states are also a common features in string compactifications, as they
arise e.g. at brane intersections, at orbifold singularities or as a result
of (brane) magnetic fluxes. In a variety of cases the only charged
string states that could propagate in the bulk are heavy (compared to the 
string scale), so that exponentially good global symmetries would arise
in the effective 4D theory as argued in the previous section.

For definiteness, in this section we will discuss the case of intersecting $D$-brane models in type-II string theories.
%
%The last question we want to answer now is whether the extradimensional axion
%construction outlined in the previous section could be uplifted to string theory.
%After all string theory provides both the natural framework where extra-dimensional gauge  theories can be UV completed and the only available theory where the quantum gravity effects we discussed can actually be computed. 
%
%We will proceed following similar steps as in section~\ref{sec:}, showing first
%how to realize the basic ingredients (PGB with high level of protection) and 
%then how to incorporate some of the other features required by the QCD axion itself.
%
Typical realistic compactifications are expected to involve rich manifolds
(such as Calabi-Yau's that can accommodate low energy supersymmetry) and an intricate web of $D$-brane configurations, leading to the SM fields and any other
sector required by phenomenology and consistency.

%%%%%%%%%%%%%%%%%%%%%%%%%%%%%%%
\begin{figure}[t!]
        \centering
         \includegraphics[width=0.7\textwidth]{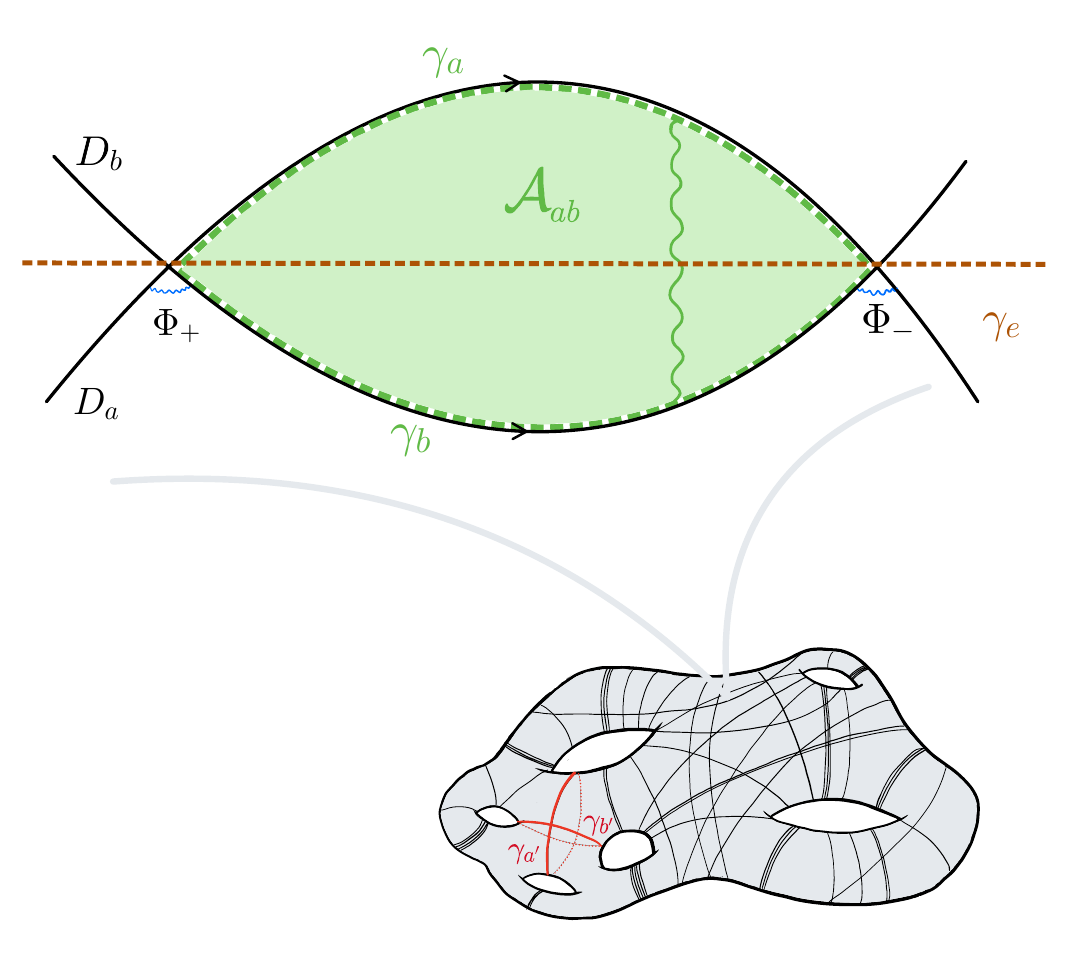}
         %
         %%%%%%%%%%% Caption %%%%%%%%%%
        \caption{In a complex compactification, two $D$ branes self-intersect twice. 
        Open strings on each intersection produce localized fields with opposite charges. An approximate global symmetry emerges under which only non-local gauge invariant operators spanning the green surface $\Sigma$ of area $\mathcal{A}_{ab}$ are charged. The symmetry can be broken only by worldsheet instantons extending in the green area or Euclidean $D$-brane instantons wrapping cycles as $\gamma_e$. Both effects are naturally exponentially suppressed. See text for more details. }
        %%%%%%%%%%%%%%%%%%%%%%%%%%%%%%%
        %
        \label{fig:branesAB}
\end{figure}
%%%%%%%%%%%%%%%%%%%%%%%%%%%%%%%
Consider two $D$ branes ($a$ and $b$) intersecting in the extra dimensions as in fig.~\ref{fig:branesAB}. Each $D$ brane is known to host a $U(1)$ gauge boson. Open strings starting and ending on the same brane will describe neutral states. The only light charged fields live at the two intersections (the strings stretching among the two $D$ branes). Focusing for the moment on the spin-0 sector we have two
complex scalar fields $\Phi_\pm$ with charges $(\pm1,\mp1)$ w.r.t. $U(1)_a\times U(1)_b$.
The charges at the two intersection are exactly opposite to each other since the intersections
in this construction are not topological (they can be removed by continuously deforming the $D$-branes, corresponding to giving a large mass to the pair). 
%The reader might worry about the stability of the construction, there are however known ways (which we will discuss later) to make such configuration at least metastable, this is indeed what happens also in typical Standard Model embeddings in intersecting brane constructions after the Higgs develops  its vev and the SM fermions become massive.

Note that none of the fields living at the two intersections is charged under the diagonal $U(1)_{a+b}$, which therefore decouple from
the rest of the system. With respect to the other $U(1)_{a-b}$, $\Phi_{\pm}$ have opposite charges. The system is therefore analogous to the one discussed in section~\ref{sec:AxionsFromWilsonLines}.
If both $\Phi_{\pm}$ develop a vev, one linear combination of the phases will become
the longitudinal component of the $U(1)_{a-b}$ vector field, while the other will serve
as a NGB. Similarly to what discussed in the 5D construction, the gauge-invariant
operator hosting the Nambu-Goldstone field would be non-local
\begin{equation} \label{eq:Wstrings}
{\cal W}_{+-}=\Phi_+\ e^{i g_a\int_{\gamma_a}\!\! A_a}\ e^{-i g_b \int_{\gamma_b}\!\! A_b}\ \Phi_-\,,
\end{equation}
and involve Wilson lines along each oriented $D$-brane path ($\gamma_{a,b}$). 
An effective potential for the NGB can only
be generated in the presence of charged fields that can connect the two intersections.
As mentioned before, there is no light charged field in this construction. There are however
heavy charged states. One is represented by open strings stretching between the two $D$-branes and traveling from one intersection to the other. 
The contribution from these states will then be suppressed by the Euclidean worldsheet action of an open string stretching the surface $\Sigma$ with area ${\cal A}_{ab}$
in between the two $D$-branes and their intersections, namely $e^{-{\cal A}_{ab}}$. The presence of such contribution can
more elegantly be argued by looking at the gauge transformation properties of the bulk NSNS 2-form field, $B_{\mu\nu}$. Under a gauge transformation, the latter shifts by $\delta B_{\mu\nu}=\partial_{[\mu}\Lambda_{\nu]}$ while $D$-brane gauge vectors shift by $g_i\delta A_{i,\mu}=\Lambda_\mu$. It follows that ${\cal W}_{+-}$ is not invariant under the gauge transformations of the field $B_{\mu\nu}$,  but needs to be appropriately dressed by the factor $e^{-i \int_{\Sigma}\hspace{-2pt} B}$. 
%where $\Sigma$ is a surface
%\footnote{Here we use the same symbol ${\cal A}_{ab}$ to indicate the Euclidean string worldsheet stretched between the $D$-branes and the value of its area.} 
%with boundaries $\gamma_{a}\cup \gamma_b$. 
The factor transforms exactly to compensate the gauge non-invariance of eq.~(\ref{eq:Wstrings}), so that the fully gauge-invariant non-local operator would now read
\begin{equation}
    \label{eq:Wstrings2}
{\cal W}_{+-}=\Phi_+\ e^{i g_a\!\int_{\gamma_a}\hspace{-5pt} A_a-i g_b\! \int_{\gamma_b}\hspace{-5pt} A_b-i\! \int_{\Sigma}\hspace{-2pt} B}\ \Phi_-\,.
\end{equation}

Parametrically ${\cal A}_{ab} \propto M_s^2 d_{ab} d_{\pm} $, where $M_s^2$ is the string tension, $d_{ab}$ is the average distance between the two $D$ branes and $d_\pm$ the distance between the two intersections. After noticing that
$M=M_s^2 d_{ab}$ is the average mass of the open string charged states that can propagate between the two $D$-brane intersections, the effect suppressed by $e^{-{\cal A}_{ab}}=e^{-M d_{\pm}}$ can be identified as the string theory version of the one in eq.~(\ref{eq:PQbreaking}). Note however
that $d_{ab} M_s$ can be parametrically large so that all charged states masses are larger than the extra-dimensional field theory cut-off. In such a case this source of breaking of the approximate global symmetry becomes harmless.

In string theory however, it is possible to generate other non-local gauge-invariant operators
involving the product $\Phi_+\Phi_-$. The RR bulk gauge fields $C$, under which
the $D$ branes are charged, are themselves charged with respect to the $D$-brane localized gauge fields. In particular, under the generic gauge transformation of the $D$-brane vector fields 
$\delta A^i_\mu=\partial_\mu \Lambda^i$, RR fields transform as $\delta C=\sum_i \Lambda^i[\pi_i]$, where $[\pi^i]$ is the form dual to the cycle $\pi_i$ wrapped by the $D_{i}$ brane.
The non local operator
\begin{equation}
 {\cal W}'_{+-}=\Phi_+\ e^{i\int_{\gamma_e}\!\! C}\ \Phi_-\,,
 \end{equation} 
where $\gamma_e$ is a closed cycle passing through both $D_a$ and $D_b$ brane intersections 
as in fig.~\ref{fig:branesAB}, would then be gauge invariant. 
Indeed the variation of the exponent is 
$i\int_{\gamma_e}\!\! \delta C=i\sum_j\int [\gamma_e]\wedge [\pi_j]\Lambda^j
=i\sum_j \int_{\gamma_e \cap \pi_j}\!\! \Lambda^j$,   providing the right phases to compensate for the variation of $\Phi_+ \Phi_-$. The non-local
operator ${\cal W}'_{+-}$ originates from Euclidean $D$ brane instantons \cite{Becker:1995kb} wrapping $\gamma_e$ and therefore it will come weighted by the exponential of the Euclidean action, i.e. $e^{-V_e/g_{str}}$, where
$V_e$ is the Euclidean $D$-brane volume (in string units) and $g_{str}$ the string coupling. At small string coupling and large volumes the contribution is again exponentially suppressed. It is maximized
when the Eucliean $D$ brane minimizes its volume. If this happens when $\gamma_e$ overlaps with one
of the $D_{a,b}$ branes, then its volume can be related to the $D$ brane gauge coupling. The
exponential suppression in that case assumes the suggestive form $e^{-8\pi^2/g_{a,b}^2}$, 
i.e. the contribution would match the one from the would-be small-instantons of the
gauge theory. Further suppression could be present if $\gamma_e$ intersects other $D$ branes, as in such a case fermionic chiral zero modes arising from
the intersections will have to be saturated in order to get a contribution for the effective scalar potential (exactly as for the usual 4D instantons). 

More general configurations can be constructed combining both string worldsheets and Euclidean $D$ branes by considering $\gamma_e$ off the $D_{a,b}$-brane intersections (an explicit example given below).  The Euclidean $D$ branes will play the role of the small instantons in gauge theories, while the string worldsheets the one of the Yukawas present in the prefactor in order to saturate any zero mode from the fermionic determinant. 
In any case, as long as the localized fields are well separated (in units of the string length), all symmetry breaking effects will be exponentially suppressed. This is in direct analogy to axions arising purely from closed string gauge fields. The level of protection of the associated NGB are therefore equivalent.

%%%%%%%%%%%%%%%%%%%%%%%%%%%%%%%
\begin{figure}[t!]
        \centering
         \includegraphics[width=0.7\textwidth]{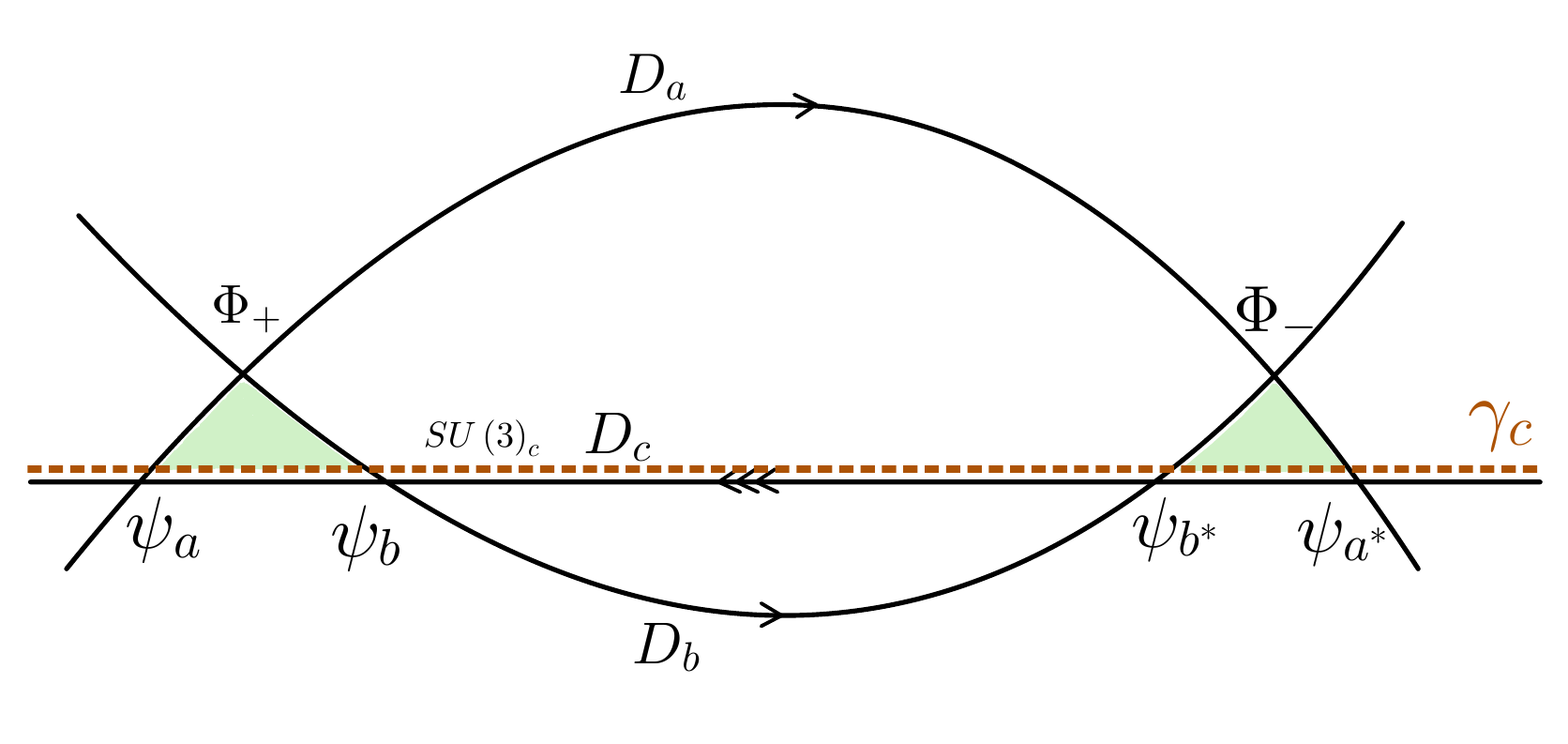}
         %
         %%%%%%%%%%% Caption %%%%%%%%%%
        \caption{When a stack of 3 $D$ branes wrapping the cycle $\gamma_c$ is added to the configuration in fig.~\ref{fig:branesAB}, charged colored states appear in the new intersections, reproducing the matter content of the model in fig.~\ref{fig:5DorbifoldSketch}. Euclidean $D$ brane instantons wrapping the same $\gamma_c$ cycle, together with worldsheet instantons extending on the green areas, produce a PQ breaking effect analogous to the one from small gauge instantons in 4D.}
        %%%%%%%%%%%%%%%%%%%%%%%%%%%%%%%
        %
        \label{fig:branesSU3}
\end{figure}
%%%%%%%%%%%%%%%%%%%%%%%%%%%%%%%
Extending the configuration to include strong interactions and have the NGB play the role of QCD axion is readily done by simply adding a stack of 3 $D$-branes as in fig.~\ref{fig:branesSU3}. The spectrum of light fields living at the 6 intersections is now 
\begin{equation}
\begin{array}{l l c l l}
\Phi_+ & [1,1,-1] & ~\qquad~ & \Phi_- & [1,-1,1] \\
\psi_a & [3,-1,0] & ~\qquad~ & \psi_{a^\star} & [\bar 3,1,0] \\
\psi_b & [\bar 3,0,1] & ~\qquad~ & \psi_{b^\star} & [3,0,-1]   
\end{array}
\end{equation}
where the quantum numbers refer to the $SU(3)$, $U(1)_a$, $U(1)_b$ respectively and we 
focused on the scalars at the $D_a$-$D_b$ intersections and on (left-handed chiral) fermions from the intersections with the $SU(3)$ brane (in supersymmetric constructions the corresponding chiral supermultiplet partners will also  appear but they do not play a role in our discussion).
Worldsheet instantons analogous to those leading to eq.~(\ref{eq:Wstrings}) produce Yukawa terms \cite{Aldazabal:2000cn} (${\rm y}_+\Phi_+ \psi_a \psi_b$ and ${\rm y}_-\Phi_- \psi_{a^\star} \psi_{b^\star}$) that could be ${\cal O}(1)$ or exponentially suppressed depending on the relative distance among the relevant intersections in string units. As before $U(1)_{a+b}$ does not play a role here\footnote{It corresponds to gauging
a baryon-like symmetry, which in more complete model is expected to be broken and decouple.}, while $U(1)_{a-b}$ and the rest of the fields reproduce our 5D example of section~\ref{sec:AxionsFromWilsonLines} 
\footnote{While the $U(1)_{a-b}$ charge assignments are slightly different from those of the model in \cref{sec:5DOrbifold} there is no qualitative difference in the way the mechanism works.}.
The string construction outlined here shows how it is possible to realize our 5D construction
in a more realistic quantum gravity uplift and more importantly justifies our assumptions
about the absence of light charged states that could generate dangerous shift symmetry breaking contributions. 
If the $SU(3)$ branes are those with the smallest worldvolume,
then $g_{a,b}\ll g_s$, PQ breaking contributions to the axion potential from worldsheet instantons would then be exponentially suppressed compared to the QCD one. The leading contribution to the axion potential beyond the IR QCD one would then come from Euclidean $D$-brane instantons. Those obtained from Euclidean $D$-branes wrapping the same compact worldvolume of the $SU(3)$ brane would be suppressed by $e^{-8\pi^2/g_s^2}$ from the Euclidean brane action and by the product of all the Yukawa couplings required to kill the fermion zero modes from each of the colored fermions. These match
the small instanton limit contribution of gauge instantons, which are generically parametrically suppressed w.r.t. the calculable contribution from low-energy QCD  and likely to be aligned with it. Other
contributions could come from Euclidean $D$-brane instantons away from the $SU(3)$ brane (passing possibly close to $\Phi_{\pm}$ intersections), they are also exponentially suppressed and subleading as long as the volume of the $SU(3)$ brane is the smallest. 

One may wonder how generic (or simple to realize) is the $D$-brane 
configuration above, given in particular that the intersections considered
are not protected by topological constraints. In fact we would
expect realistic compactifications to be complex enough to host non topological 
intersections such as those from $D$-branes bended by the presence of fluxes/curvature or wrapping metastable cycles (as e.g. $\gamma_{a'}$ and $\gamma_{b'}$ in fig.~\ref{fig:branesAB}). Another simple way to realize the setup is from $D$-brane recombination
in a configuration of straight $D$ branes and corresponds to the Higgsing of localized fields at some intersection (similarly to what happens in the Standard Model from intersecting $D$-brane compactifications \cite{Aldazabal:2000cn,Cremades:2003qj}). This last construction could in principle be implemented 
starting from a supersymmetric $D$-brane configuration, 
in this way all the generated vevs (both for the fields associated to the brane recombination and for the PQ fields) can be kept parametrically
 small, typically of order the SUSY soft terms,  decoupling them from the string scale. We will give an explicit example of such string construction in appendix~\ref{app:susystring}.   

In the absence of low scale SUSY (which is however propaedeutic to stable string compactifications), we should expect large mass terms for the localized scalars, hence 
the vevs would not be four dimensional. 
High-quality symmetries would 
however still be present in 4D and manifest themselves as
chiral symmetries of the light fermion fields from intersections. Non-Abelian branes intersecting Abelian
branes multiple times would produce non-Abelian gauge theories
with almost massless vectorlike quarks. Upon confinement, 
light pions emerge, some of them potentially serving as  exponentially light composite axions.

 %%%%%%%%%%%%%%%%%%%%%%%%%%%%%%%%%%%%%%%%%%%%%%%%%%%%%%%%%%%
\subsection*{Opening the string axiverse}
%\label{sec:StringTheory2}

Zooming out from the specific construction described above for the QCD axion,
a much richer structure emerges.  
Many string theory compactifications, such as Calabi-Yau's 
that preserve supersymmetry to some degree, possess a large 
number of non-contractible cycles. Zero modes of higher rank
gauge fields wrapping such cycles could potentially produce 
a multitude of high-quality axion-like particles (ALPs), the
collection of which is sometimes referred to as the {\it string
axiverse} \cite{Arvanitaki:2009fg}. In a similar way, phenomenologically realistic type II
string compactifications possess a rich structure 
of $D$ branes. Each of them generically have multiple intersections,
leading to a plenitude of localized charged fields with exponentially good
emergent global symmetries, as argued above.
It is therefore not unreasonable to expect that a number of them
could be in the spontaneously broken phase. This could happen
as described in the example above as a result of localized charged 
scalars getting a small vev (e.g. as a result of SUSY breaking) or,
even in the absence of SUSY, from chiral fermion condensation 
after confinement of non-Abelian sectors with exponentially light quarks.
This leads to what we would call an {\it open string axiverse}. The main qualitative difference among the closed and the open axiverses is that in the latter the axion decay constants are fully four dimensional and parametrically separated from the string scale (e.g. if the SUSY breaking  scale is below the
string scale or if the spontaneous breaking arises as a result of dimensional transmutation). 

This, in turn, opens the possibility of a calculable post-inflationary axiverse with important phenomenological implications, such as a more predictive target parameter space for ALP dark matter, gravitational signals associated with topological defects, and potential signatures from small-scale dark matter substructures, such as axion mini-halos and Bose stars, as we will discuss in the next section.

%%%%%%%%%%%%%%%%%%%%%%%%%%%%%%%%%%%%%%%%%%%%%%%%%%%%%%%%%%%
\section{Cosmology}
\label{sec:cosmology}
%%%%%%%%%%%%%%%%%%%%%%%%%%%%%%%%%%%%%%%%%%%%%%%%%%%%%%%%%%%

% The cosmic history of axions depends firstly on whether or not their associated PQ symmetry is restored after inflation. If not, then  

Our stated objective in generalizing the basic mechanism of quality protection from extra dimensions was to make this compatible with the more predictive post-inflationary scenario. 
This holds whenever the universe starts with a maximum temperature $T_{\rm max}$ (alternatively, a scale of inflation) exceeding the PQ phase transition scale, at which point PQ symmetry is spontaneously broken and the effectively massless axion becomes a well-defined degree of freedom\footnote{This is to be contrasted with the `pre-inflationary' scenario, where an initial, `misaligned', constant field value $\langle a/v \rangle =\theta_{\rm i}$ is an incalculable extra parameter that goes into the axion dark matter abundance $\Omega_{{\rm mis}, \, \theta_{\rm i}}(m,v)$.
}.
In the process, the axion field takes on random values $a/v\in (0,2\pi)$, initially uncorrelated on scales set by the phase transition temperature, and in this manner cosmic strings (see ref.~\cite{Vilenkin:2000jqa} for a review), topologically stable solutions to the field equations, form by the well-known Kibble mechanism \cite{Kibble:1976sj,Kibble:1980mv}.

% For axions arising from the closed string sector of the original axiverse, axion strings are profiles ultimately sourced at their core by magnetically charged $D$ branes \cite{Benabou:2023npn,March-Russell:2021zfq}. The only set ups proposed to produce these non-perturbative string theory objects consistent with our cosmology are poorly understood, such as $D$-brane inflation and a Hagerdon phase transition \cite{Benabou:2023npn,xxx} .

% For axions arising from the closed string sector of the original axiverse, PQ symmetry is never restored. Rather axion strings are profiles ultimately sourced at their core by magnetically charged $D$ branes \cite{Benabou:2023npn,March-Russell:2021zfq}. The only set ups proposed to produce these non-perturbative string theory objects consistent with our cosmology are poorly understood, such as $D$-brane inflation and a Hagerdon phase transition \cite{Benabou:2023npn,xxx} . \flag{xxxxx}

% In the constructions of \cref{sec:AxionsFromWilsonLines} there are three relevant scales: the KK scale $L^{-1}$ and the brane-localized scales of spontaneous symmetry breaking $v_\pm$, with $v$ parametrically set by the smallest as per \cref{eq:AxionCombinationOrbifold}. 
In the constructions of \cref{sec:AxionsFromWilsonLines,sec:StringTheory}, the 
axion-defining line operators, such as \cref{eq:WilsonLine}, serve as an order parameter for the phase transition, with $\langle |\mathcal{W}| \rangle = v_+v_-/2$
in the broken phase, and $\langle |\mathcal{W}| \rangle = 0$ can be straightforwardly restored for temperatures $T\gtrsim {\rm min}(v_-,v_+)$. 
We will take $v_- \leq v_+$ without loss of generality, and will always consider $v_-\ll L^{-1}$ to avoid issues of stability of the radion at temperatures above the KK scale. 
We start by highlighting the \textit{KK}SVZ-like regime $v_- \lesssim T_{\rm max} \ll v_+,L^{-1}$, in which we recover the minimal post-inflationary scenario realized in 4D field theory. We discuss the interesting but more complicated case of $T_{\rm max} \gtrsim v_+$ in \cref{sec:multiString}.
Finally, we note that single windings of $\theta_-=\arg(\Phi_-)$, result in single windings of the fundamental domain of $a/v$, as can be seen from the definition $a/v \equiv \arg(\cal W)=\theta_- + \dots$, as $\cal W$ contains only one power of $\Phi_-$ in all the examples discussed\footnote{This is not the case instead if $|q_-|/|q_+|\not \in \mathbb{N}$, as discussed in a purely 4D field-theoretic context in ref.~\cite{Lu:2023ayc}.}.

% We note that in general, the number of windings $n$ in the (gauge-invariant) fundamental domain $2\pi v$ around the strings dominantly produced during the phase transition is a dynamical and theory-dependent question \cite{Lu:2023ayc}.  For the cases highlighted in this work $n=1$. This is easily seen from \cref{eq:AxionDef} or \cref{eq:AxionOrbifold}, by the fact that single winding of $\theta_-=\arg(\Phi_-)$, which will statistically be the most dominant occurrence as $T\lesssim v_-$, results in single winding of $a/v$.

A network of cosmic strings eventually evolves towards an attractor `scaling' solution \cite{Kibble:1976sj,Kibble:1980mv,PhysRevD.24.2082}, with energy density $\rho_s =  \xi \mu/ t^2$. Here $\xi$ is the string length per Hubble patch, with mild time dependence $4\xi \approx \log(m_\rho/H(t))$ suggested by numerical extrapolation (e.g. see ref.~\cite{Gorghetto:2020qws}), and $\mu \approx \pi v^2\log\left(m_\rho/H(t)\right)$ is the typical tension of a global (axion) string.
% , where the IR divergence is regulated by the typical separation between strings $\sim H^{-1}$.
We take the mass of the radial mode $m_\rho \sim v$ unless stated otherwise.
% $\rho_s$ does not include the constantly forming sub-horizon loops, whose dynamics and lifetime can in general be more complicated and can be a source of uncertainty, depending on the observable of interest. 
During this regime, strings convert an $\mathcal{O}(1)$ fraction of their energy density into axions per Hubble time, i.e. with rate $\Gamma_s \approx \rho_s/t$. 
Scaling continues until Hubble $H(t)$ drops below the scale $H_* \equiv m_a(T_*)$, the mass of the axion, after which the configuration becomes sensitive to the axion potential. 
For the QCD axion, this derives from \cref{eq:QCDaxion_defining_interaction} with periodicity $V_{\rm QCD}(a) = V_{\rm QCD}(a+2\pi v/N_w)$, and 
$N_w$ domain walls branch out of each singly winded string. For $N_w=1$, such as the minimal benchmark model highlighted in this work, each domain wall starts from one piece of string and ends on another piece of opposite orientation, eventually pulling them together, unwinding them, and  the string network decays.
% \footnote{\flag{xxx} While analytic estimates exist, a precise, quantitative understanding of even this minimal post-inflationary scenario ultimately requires numerics. First-principle, field-theoretic simulations are limited by large scale separation $\log({v/m})$ in the problem, requiring some form of extrapolation.}.
% , currently unmanageable beyond $\log(f_a/H) \approx 8$ \cite{Gorghetto:2018myk}. 
% Much work has gone into extrapolating a  prediction for the final axion dark matter abundance \cite{}. 

For $N_w>1$, domain wall solutions that interpolate between two of the $N_w$ distinct and degenerate vacua are topologically stable.  In the cosmological setting, this leads to the continued survival of the network of strings plus domain walls \cite{Sikivie:1982qv}. When the latter become energetically dominant, a new scaling regime is believed to  kick in, with energy density $\rho_{\rm DW} =  \mathcal{A}_w \sigma /t$, where $\sigma = 8 m f_a^2$ is their surface tension, and $\mathcal{A}_w$ parametrizes the domain wall area per Hubble patch. $\mathcal{A}_w$ is expected to go like $N_w$, with some limited numerical validation \cite{Gorghetto:2022ikz}. This continues until breaking effects of the remaining $\mathbb{Z}_{N_w}$ symmetry become relevant. 
In the case of the QCD axion, a $\mathbb{Z}_{N_w}$-breaking tilt in the potential consistent with solving the strong CP problem and not overproducing dark matter pushes $f_a$ down towards astrophysical bounds.
% \cite{Chang:1998tb,xx}. 
For an ALP there is no constraint on the tilt and a period with domain walls is perfectly allowed, granted they decay early enough to be consistent with cosmological observations. In analogy with \cref{eq:QCDaxion_defining_interaction}, we define $f\equiv v/N_w$ for an ALP, where $\mathbb{Z}_{N_w}$ is now the discrete symmetry preserved by whatever contribution gives the largest $U(1)$ breaking and thus determines the ALP's mass.

% This is to be compared with misalignment, and no post-inflationary dynamics.

%  

% Another difference compared to the usual extra-dimensional axions, is that the decay constant $f_a$ in our setup is independent from the KK scale although a natural separation $f_a \ll L^{-1}$ calls for supersymmetry of composite dynamics, as explored in \cref{sec:Naturalness}.

%%%%%%%%%%%%%%%%%%%%%%%%%%%%%%%
\begin{figure}[t]
        \centering
         \includegraphics[width=0.9\textwidth]{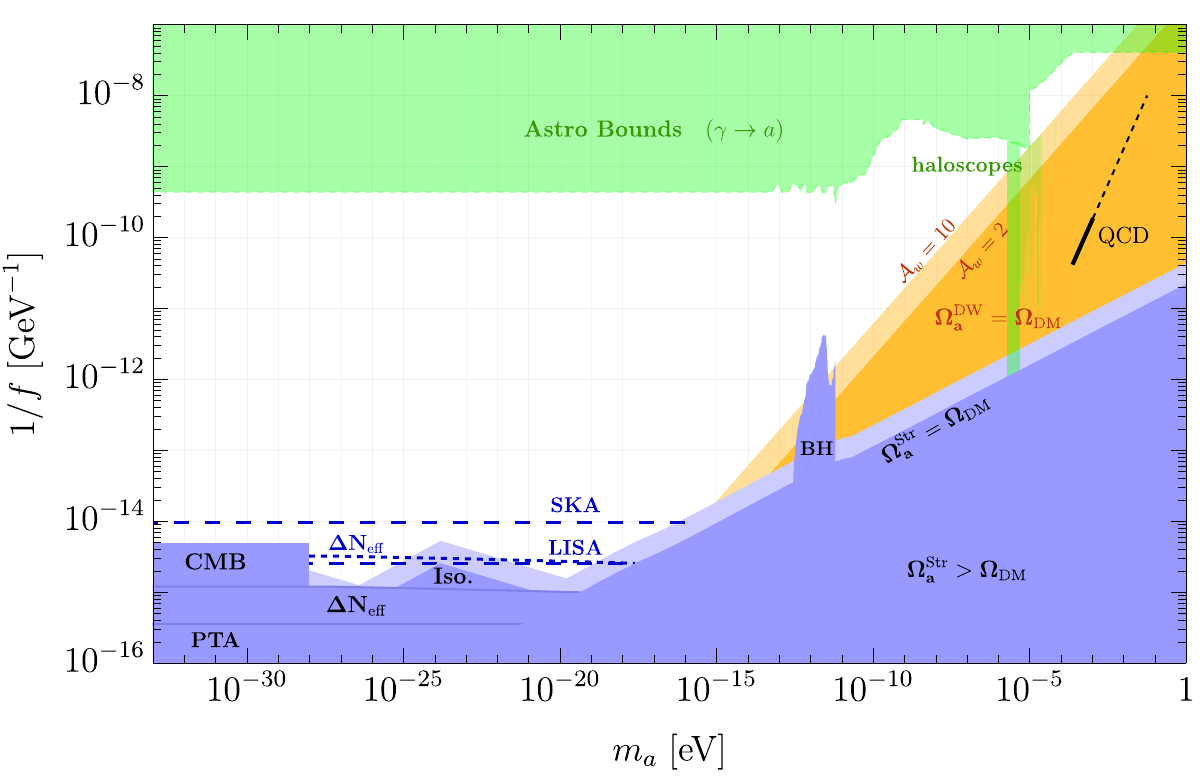}
         %
         %%%%%%%%%%% Caption %%%%%%%%%%
        \caption{Constraints on a minimal post-inflationary ALP scenario with constant mass $m$ and decay constant $f$. See \cref{sec:postInflationaryALPs} for all details.
        Blue:  Several gravitational constraints at large $f$, most unique to the post-inflationary scenario, resulting from dynamics of large tension topological defects ($N_w=1$). Blue dashed indicates the sensitivity of some future experiments.
        Orange: In the case with metastable domain walls ($N_w>1$), the region highlights where axions from their decay can make up the entirety of dark matter consistent with estimated isocurvature bounds. 
        Green: Assuming a coupling to photons, astrophysical bounds and haloscope searches rule out lower values of $f$. 
        % We highlight bounds from CAST \cite{CAST:2024eil} and the Chandra Observatory \cite{Reynes:2021bpe}.
        % \cite{Ramberg:2019dgi} study GWs from global strings in non-standard cosmologis and I refuse to cite them.
        }
        %%%%%%%%%%%%%%%%%%%%%%%%%%%%%%%
        %
        \label{fig:ALPplot}
\end{figure}
%%%%%%%%%%%%%%%%%%%%%%%%%%%%%%%

%%%%%%%%%%%%%%%%%%%%%%%%%%%%%%%
\subsection{Post-inflationary ALPs}
\label{sec:postInflationaryALPs}
%%%%%%%%%%%%%%%%%%%%%%%%%%%%%%%

Having qualitatively described the dynamics of an axion in a post-inflationary scenario, in this subsection we discuss in some detail constraints and some detection opportunities on their parameter space, motivated by the plausible existence of a post-inflationary axiverse, as argued at the end of \cref{sec:StringTheory}. 
We explicitly consider here only the most minimal and predictive implementation, assuming a single axion and standard cosmology.
The ALP has then an in principle predictable cosmic history determined by its decay constant $f$ and mass $m$, which we now take to be time-independent, as typical in the constructions of \cref{sec:StringTheory}. In \cref{fig:ALPplot} we take a broad view, spanning masses as small as $H_0$ (with the string network surviving till the present day) and up to $1 \eV$. 
% Everywhere in this plot $f\leq v\lesssim T_{\rm max}$ by assumption. 
This parameter space has been studied already in refs.~\cite{Gorghetto:2021fsn,Gorghetto:2022ikz}, and all our findings are compatible therewith. 
We warn the reader that most boundaries in the figure are only indicative as they are sensitive to parameters which are either extrapolated from current simulations, with still significant uncertainties, or obtained from educated guesses when these are not available. We first give a summary description of \cref{fig:ALPplot}, leaving quantitative details to dedicated paragraphs further down.

% \flag{We warn the reader that most of the boundaries of the allowed parameter space discussed in the following are only indicative as they are sensitive to parameters which are either extrapolated from current simulations with still significant uncertainties or obtained from educated guesses when numerical simulations are not available. 
% Also we explicitly discuss here only the most minimal and predictive implementation of the open string axiverse, assuming, for instance, that at most one axion dominate the dark matter abundance, no important level-crossing effect from multiple axions is at work, standard cosmology, etc. Much richer (though less predictive) phenomenology is possible if these assumptions are relaxed as already partially discussed
% in the literature of the closed string axiverse.}

The observed dark matter abundance $\Omega_{\rm DM}\simeq 0.26$ is obtained in principle on a definite line ($\Omega^{\rm Str}_a=\Omega_{\rm DM}$), which could serve as a target for future detection (see for example haloscope projections in ref.~\cite{Gorghetto:2022ikz}). 
In practice, uncertainty remains in the theoretical prediction, which we 
represent by the light blue band,
% and we draw two lines, one crudely estimating the abundance from final network collapse  $(\Omega_{\rm UV})$, the other assuming an IR dominated spectrum of axion emission from strings $(\Omega_{\rm IR})$. 
details of which are discussed further down. The region below the line is excluded by overproduction of dark matter, while above the ALP forms a decreasing fraction of it. 

A caveat is when $N_w>1$ and domain walls boost the axion abundance by an amount controlled by an extra parameter --- the size of $\mathbb{Z}_{N_w}$ breaking. In this case, post-inflationary axions can make up the totality of dark matter, consistent with constraints studied here, everywhere in the orange region labeled by $\Omega_a^{\rm DW} = \Omega_{\rm DM}$, as quantified further down. Till then, we continue assuming $N_w=1$.

For comparison, we also display the QCD axion line towards the top-right of the figure. The observed dark matter abundance in principle is now a point. The uncertainty in the theoretical prediction is reflected by the black line's finite length of $0.25 \lesssim m_a/\meV \lesssim 1$, as per the extrapolation in ref.~\cite{Gorghetto:2020qws,Kim:2024wku}, compatible also with results of ref.~\cite{Benabou:2024msj}.
The line turns dashed when the abundance is less than $ \Omega_{\rm DM}$ and stops when $f_a \lesssim 10^{8}\GeV$ from the SN1987A cooling bound \cite{Raffelt:2006cw} on \cref{eq:QCDaxion_defining_interaction}.

% \footnote{If $g_{a\gamma\gamma}=0$, then \cref{fig:ALPplot} could be continued to far smaller values of $f$. We leave the study of the photon-coupling-independent (and likely UV-model-dependent) bounds in this limit to future work.} 
The sharply peaked black hole superradiance bounds around $m\sim 10^{-12}\eV$ from observations of highly spinning stellar mass  black holes \cite{Baryakhtar:2020gao,Witte:2024drg,Hoof:2024quk} are independent of cosmological history.
% (less robust bounds exist also at $m\lesssim 10^{-18}\eV$ from supermassive black holes \cite{Hoof:2024quk,Mehta:2020kwu}). 
By contrast to its pre-inflationary cousin, the post-inflationary scenario also comes with further purely gravitational constraints and possible signals at large $f$, due to the presence and dynamics of topological defects, whose tension increases with $f$. 
For very light masses $m\lesssim 10^{-28}\eV$, the string network persists at the epoch of decoupling and is constrained directly by the lack of observed string-like discontinuities in temperature fluctuations in the CMB
from Planck data \cite{Planck:2013mgr,Planck:2015fie,Lopez-Eiguren:2017dmc}. This bounds the string tension, $G\mu \lesssim 10^{-7}$ very roughly, where $G$ is Newton's constant, resulting in $f\lesssim 5 \cdot 10^{15}$. 
Potentially strong bounds, though difficult to quantify precisely, come from structure formation, comparing isocurvature matter density fluctuations in the component of dark matter made of post-inflationary axions with observations consistent with an initially quasi scale-invariant adiabatic spectrum, as detailed below. These are the peaks labeled `Iso' in \cref{fig:ALPplot}. They inherit the uncertainty of the dark matter abundance prediction (again, light blue), as well as that on the spectrum, which we do not quantify.
% \flag{xxx}
% For $N_w>1$, assuming they produce the totality of $\Omega_{\rm DM}$, they are the appropriately labeled lightly shaded band overlaying the blue region of $\Omega_{\rm DW} \leq \Omega_{\rm DM}$.
Relativistic axions emitted by the strings will contribute extra free-streaming dark radiation.
We show the current bound and projected reach (blue dashed) of near-future experiments, as quantified below. 
Finally, gravitational waves emitted by the string network contribute to a potentially observable stochastic background \cite{Chang:2019mza,Gorghetto:2021fsn,Chang:2021afa}. The current constraint $f_a \lesssim 2.8  \times 10^{15} \GeV$  is taken from ref.~\cite{Servant:2023mwt} using the 15-year NANOGrav data.
Indicative projected sensitivities (blue dashed) displayed for future gravitational wave detectors (LISA \& SKA) are taken directly  from ref.~\cite{Gorghetto:2021fsn}.

\textit{Assuming} a coupling to photons $g_{a\gamma \gamma} \equiv {\alpha_{\rm em}\over 2 \pi f}c_{a,\gamma}$, several exclusion bounds on low $f$ follow from astrophysics (regardless of cosmology), as well as from haloscope cavities looking for axion dark matter today.
These are plotted in green for $c_{a,\gamma}=1$ using combined data from ref.~\cite{AxionLimits}. The ADMX haloscope band \cite{ADMX:2024xbv}  reaching down to the $\Omega^{\rm Str}_a=\Omega_{\rm DM}$ boundary applies both for $N_w=1$ and for $N_w>1$, while the fainter haloscope bounds to its right apply only for the orange region that assumes $\Omega_a^{\rm DW} = \Omega_{\rm DM}$.

\paragraph{Dark matter abundance:}
The final abundance of axion dark matter can be roughly separated into two main contributions: axion emission during the scaling regime up to $H=H_*$, and the more complex final decay of the network induced by ($N_w=1$) domain walls.
The first depends on accurate knowledge of $\xi(t)$, and the spectrum of energies 
emitted\footnote{The only scales in the problem during scaling are $f$ and $H(t)$. At these extremes, axions produced with energy $\sim f$ remain relativistic to the present day, while those emitted with energy $\sim H(t)$ have an energy of $\sim \sqrt{H(t) m}$ when the string network collapses and can contribute to dark matter.}.
We use the semi-analytic expression derived in ref.~\cite{Gorghetto:2020qws} for it,
specifically eq.~(36) of Appendix C therein, which assumes an IR-dominated instantaneous emission spectrum at late times. Simplified for present purposes, this gives
\begin{align}
  \begin{split}
        \Omega^{\rm Str}_{a}  &\approx  (4\pi \xi_* \log(f/m))^{3/4}  \Omega_{\rm mis, \, \theta_{\rm i}=1}  \\
        & \simeq 0.23  \left( (4\pi \xi_* \log(f/m)) \over 1.3\cdot 10^4 \right)^{3\over 4}\left({m\over  10^{-7} \eV }\right)^{1 \over 2} \left( f \over 10^{12} \GeV \right)^2 \left({ 90 \over g_{*}(T_{*})}\right)^{1 \over 4}  
        \label{eq:OmegaStr} \ ,
  \end{split}
    %%%%%%%%%%%%%%%%%%%%%%%%%%%%%%%%%%%%%%%%%%%
    % \left(g_*(T_*) \over g_*(T_{nr})\right)^{1/12} \ .
    %%%%%%%%%%%%%%%%%%%%%%%%%%%%%%%%%%%%%%%%%%%
    % \Omega_{alp, \rm miss} 
\end{align}
where $\xi_* \equiv \xi(t_*)$, and $\Omega_{{\rm mis}, \, \theta_{\rm i}=1} $ is the ($\theta_{\rm i}=1$) misalignment prediction\footnote{Obtained by red-shifting the axion number density $\theta_{\rm i}^2 m f^2$ at $t_*$ to the present day.} for comparison. The $\Omega^{\rm Str}_{a}=\Omega_{\rm DM}$ lines in \cref{fig:ALPplot} are plotted multiplying and dividing  $\Omega^{\rm Str}_{a}$ by a factor of 2, to emphasize uncertainties in the various numerical extrapolations of ref.~\cite{Gorghetto:2020qws} and the potential enhancement from network decay.
The abundance (\ref{eq:OmegaStr}) can be parametrically understood as follows. For an IR-tilted emission spectrum, the axion number density is dominated by late emission. Thus, the relevant energy density at $t_*$ is approximately $ 4 \xi_* \mu_* H_*^2 = 4\pi \xi_* \log(f/m) f^2m^2$. Since this is much larger than the potential energy density ($V_a \approx f^2 m^2$),
these axion waves are still relativistic and redshift like radiation until the Hubble scale $H_{\rm nl} = m/\sqrt{4\pi \xi_* \log(f/m)}$
% factor is $R_{\rm nl} = R_* \left(4\pi \xi_* \log(f/m)\right)^{1/4}$
, when kinetic and potential energies approximately match\footnote{Here `nl' comes from the importance of non-linearities in the axion potential during this stage --- see ref.~\cite{Gorghetto:2020qws}}. This dilution, translated back to $t_*$, then gives the power of 3/4 in \cref{eq:OmegaStr}.

Even if the contribution from strings were significantly smaller than (\ref{eq:OmegaStr}), e.g. if the emission spectrum at late times turned out to not be IR dominated (as advocated for example in \cite{Benabou:2024msj}), the second contribution to the axion abundance from domain wall decays, is plausibly not much smaller.
We can very crudely estimate this by supposing that the domain walls, whose energy density initially grows as $\xi \sigma H$, promptly collapse at $H\approx m/\xi$, when all $\xi$ walls (of size $m^{-1}$) are properly resolved in a Hubble patch. This results in an approximate abundance  $8 \xi_*^{3/2}\Omega_{\rm mis, \, \theta_{\rm i}=1}$. 
% \langle \theta_{\rm i}^2\rangle
It is of some comfort to the theoretical prediction that this estimate is numerically within the proposed error on (\ref{eq:OmegaStr}).

% If the true spectrum if not IR-dominated, the contribution from strings could be substantially smaller, however we expect the contribution from domain walls not to much smaller. Since xi si large domain walls are resolved have to wain till when $H=m/\xi$. This delay procues enhancement of $\xi^{3/2}$ with respect one domain wall per Hubble patch. Not far from the plotted band. 

% A useful but very crude approximation for the dark matter abundance estimates the contribution from final domain wall collapse by simply taking the average over the missalignment prediction, the logic being that all angles $a/v$ are present at $H_*$. This gives
% \begin{align}
%     \label{eq:OmegaUV}
%     \Omega_{a, \rm UV} &\sim \langle \theta_{\rm i}^2 \rangle \Omega_{\rm mis, \theta_{\rm i}=1} \approx \sqrt{m\over  10^{-21} \eV} \left( f \over 10^{17} \GeV \right)^2 \left({ 4 \over g_{*}(T_{*})}\right)^{1/4}   \ ,
% \end{align}
% where $\langle \theta_{\rm i}^2 \rangle \simeq 4.6$ accounting also for anharmonicities \cite{GrillidiCortona:2015jxo} and the label `UV' will soon be explained. This however ignores possible contributions from the string network. 

\paragraph{Isocurvature fluctuations:}
Constraints are formulated in terms of the dimensionless power spectrum $\mathcal{P}(k,t)$ for matter density fluctuations $\delta = \left[\rho(\vec{x})-\bar{\rho}\right]/\bar{\rho}$, with 
($\bar{\rho}$) $\rho$  the (average) energy density,  defined by $ \langle \tilde{\delta}^*_{\vec{k}} \tilde{\delta}_{\vec{k}'} \rangle = {(2\pi)^3}\delta^3 (\vec{k}-\vec{k}')\mathcal{P}/k^3$,
% \begin{align}
%     \langle \tilde{\delta}^*_{\vec{k}} \tilde{\delta}_{\vec{k}'} \rangle = {(2\pi)^3}\delta^3 (\vec{k}-\vec{k}')\mathcal{P}/k^3\ ,
% \end{align}
with $\tilde{\delta}_{\vec{k}} = \int d^3x \delta e^{-i \vec{k}\dot \vec{x}}$ the Fourier transform. 
At matter-radiation equality, the power spectrum of 
adiabatic fluctuations is given by $\mathcal{P}_{\rm ad}(t_{\rm eq},k)=T_{\rm ad}^2\Delta_{\zeta}$, where the transfer function $T^2_{\rm ad}(t_{\rm eq},k)$ accounts for the dynamical evolution of subhorizon modes from the initial scale-invariant spectrum $\Delta_\zeta = A_s (k/k_p)^{n_s-1}$, with \( A_s = 2.2 \times 10^{-9} \), \( n_s \approx 1 \), and $k_p=0.05 \Mpc^{-1}$ the pivot scale \cite{Planck:2018vyg}. At equality, $\mathcal{P}_{\rm ad}$ goes from $\approx A_s$ around $k \approx k_p$ to $\sim 10^{-6}$ towards $k\sim k_{\rm obs} = 10 \Mpc^{-1}$, which are the shortest scales for which we have direct observations through the Lyman-alpha forest (see e.g. ref.~\cite{Irsic:2019iff} and references therein).

For post-inflationary axions, order unity fluctuations plausibly exist at scales of order  $m$ at $H_{\rm nl}$. 
For much smaller $k$ (larger distances), fluctuations are uncorrelated by causality and the dimensionless power spectrum will take on the characteristic white noise scaling $\mathcal{P} \approx  \left(\Omega_a/\Omega_{\rm DM}\right)^2(k/k_{\rm wn})^3$, where we assume $k_{\rm wn} = m R_* \left(4\pi \xi_* \log(f/m)\right)^{1/4}$. This choice is parametrically the same as in ref.~\cite{Amin:2022nlh} and is numerically similar to that in ref.~\cite{Gorghetto:2022ikz}.
% The true power spectrum is unknown, and the reader is  qualitative distortions are discussed in \cite{Gorghetto:2022ikz} 
The transfer function in this case is close to one \cite{Amin:2022nlh} and we can directly compare to $\mathcal{P}_{\rm ad}(t_{\rm eq},k)$ above.
For the largest axion masses $k_{\rm wn} \gg k_{\rm obs}$, the strongest constraint comes from comparing the power spectra at the smallest observable scales, demanding $\left(\Omega_a/\Omega_{\rm DM}\right)^2(k_{\rm obs}/k_{\rm wn})^3\lesssim 10^{-6}$ \cite{Irsic:2019iff}. Once the peak falls below this scale, $k_*<k_{\rm obs}$, we  continue to impose $\left(\Omega_a/\Omega_{\rm DM}\right)^2\lesssim 10^{-6}$ for simplicity. In a similar fashion we also impose at CMB scales $\mathcal{P}(k_p)\lesssim 10^{-9}$ \cite{Feix:2020txt}.

\paragraph{Dark radiation:}
Bounds are quoted in terms of the effective number of extra relativistic neutrino species $\Delta N_{\rm eff}\equiv 8/7(11/4)^{4/3}\rho_{\rm extra}/\rho_{\gamma}$, currently constrained at $95\%$ confidence level to be $\lesssim 0.3$ at CMB \cite{Planck:2018vyg}.
For $m\lesssim  10^{-28}\eV$, all axions emitted during the scaling regime are still relativistic at decoupling.
It is then easy to show that the integrated energy density emitted $\int^{t}dt'\Gamma'_s (a'/a(t))^4   \approx {\pi \over 3}f^2 H(t)^2\log\left(f/H(t)\right)^3$ results in a present bound of $f\lesssim 10^{15}\GeV$. The constraint changes only logarithmically as $m \gg H(t_{\rm eq})$, as already pointed out in ref.~\cite{Gorghetto:2021fsn}, since even for an IR-dominated emission spectrum, the energy density of axions at late times is approximately uniform on a log scale\footnote{See for example the spectrum eq.(23) of appendix C in \cite{Gorghetto:2020qws}, which we integrate with a lower cut-off momentum $m\sqrt{m/H_{\rm eq}}$ to draw the $\Delta N_{\rm eff}$ lines in \cref{fig:ALPplot}.}.
% \footnote{Even if one assumes, during the scaling regime, the unphysical case of \textit{all} axions instantaneously emitted with IR cut-off energy $H(t)$, the energy density in relativistic axions at $t_{\rm eq}$ is $\approx {\pi \over 3}f^2 H_{\rm eq}^2\log\left(f H_{\rm eq}/m^2\right)^3$ for $m\gg H_{\rm eq}$. Integrating a more realistic spectrum such as can be found in Appendix C of \cite{Gorghetto:2020qws}, one finds \flag{xxx}}.
The sensitivity is projected to improve to about $\Delta N_{\rm eff}\lesssim 0.04$ \cite{TopicalConvenersKNAbazajianJECarlstromATLee:2013bxd,CMB-S4:2016ple}, which would move the bound to $f\gtrsim 3\times 10^{14}\GeV$, as shown in figure.

We note that, apart from that produced from string emission, a post-inflationary scenario comes also with a minimum contribution from thermally produced axions, by the assumption of thermal equilibrium around the PQ phase transition. Assuming standard cosmology, this lower bound is
\begin{align}
    \Delta N_{\rm eff} \gtrsim 0.027 \left(T_{\rm PQ} \over T_{\rm SM, PQ}\right)^4 \left({ 106.75 \over g_{*}(T_{\rm SM, PQ})}\right)^{4 \over 3} \ , \qquad \text{for} \ m \lesssim 0.3 \eV \left(T_{\rm PQ} \over T_{\rm SM, PQ}\right) \ ,
    % \left(106.75\over g_{*,s}(T_{\rm SM, PQ})\right)^{4/3}
\end{align}
where we have allowed the temperatures of the axion and Standard Model sectors at the PQ phase transition to differ. Futuristic CMB-HD experiments \cite{Sehgal:2019ewc} currently project a sensitivity of $\sigma(N_{\rm eff})=0.014$, and might therefore probe most of \cref{fig:ALPplot} for the minimal (arguably more plausible) case of $T_{\rm SM, PQ}= T_{a,\rm PQ}$, unless $g_*$ changes significantly. In SUSY, however, $g_*$ doubles, reducing the signal down to at most $0.011$.

\paragraph{Domain walls:} 
For $N_w>1$, the energy density in domain walls eventually ends up dominating since it redshifts less than that in strings. In this case, the region of parameters space for which ALPs could account for the whole observed dark matter abundance extends to smaller values of masses and decay constants. Assuming that the domain wall network decays into ALPs with momentum of order $m$ at temperature $T_d$, the
right dark matter abundance is obtained for
\begin{equation}
\label{eq:f_from_Td}
    f\approx 10^9~{\rm GeV} \frac{g_d^{1/4}}{{\cal A}_w^{1/2}}\left(\frac{T_d}{100~{\rm keV}} \right)^{\frac12}\left(\frac{10^{-5}~{\rm eV}}{m} \right)^{\frac12}
\end{equation}
where $g_d$ are the number of dof in thermal equilibrium at $T=T_d$.
To be clear, we are assuming that there exists an appropriate amount of $\mathbb{Z}_{N_w}$ breaking, collapsing the network at $T=T_d$. 
Bounds from structure formation, following the same logic as above, will constrain $T_d \gg \eV$, well before matter-radiation equality.
We assume that the white noise tail for the power spectrum, in this case, is
$\mathcal{P}(k) = (k/k_{\rm DW})^3$, with $k_{\rm DW}={\cal A}^{1/3}_w H_d R_d$, set by the Hubble scale at decay.
% i.e. we assume ${\cal O}(1)$ density fluctuations on the scale of the average distance between domain walls at decay. 
This assumption is only an educated guess in the
absence of a dedicated study.  Demanding $\mathcal{P}(k_{\rm obs})<10^{-6}$ 
leads to $T_d \gtrsim 0.1  \MeV /\mathcal{A}^{1/3}_w$. Plugging this into \cref{eq:f_from_Td} defines the boundaries of the orange region in \cref{fig:ALPplot}.

\subsection*{Small scale structures}
The isocurvature density fluctuations arising from the decay of strings and domain wall networks, which we used to put bounds on part of the parameter space, are also 
known to produce self-gravitating small scale structures, usually known as axion miniclusters or mini-haloes \cite{Hogan:1988mp,Kolb:1993zz}. 
For the ($N_w=1$)  QCD axion, a recent study \cite{Gorghetto:2024vnp} found that structure forms promptly at matter-radiation equality, primarily as Bose stars, i.e. their typical size matches the de~Broglie wavelength of the component axions. This is not expected for ALPs with a temperature-independent mass \cite{Gorghetto:2024vnp}.
For $N_w=1$, the quantum Jeans scale at equality $k_J\sim \sqrt{H_{\rm eq} m}\, R_{\rm eq}$ is smaller than  the scale $k_{\rm wn}$ at which $\mathcal{P}$ is ${\cal O}(1)$, which inhibits the formation
 of structure at equality (only structure with $k < q_J \sim k_J^2/k_{\rm wn}< k_J$ can grow, $q_J$ being the classical Jeans scale $q_J=\sqrt{4\pi G\rho_{\rm eq} }R_{\rm eq}/u_{s}^{\rm eq} $, and $u_{s}^{\rm eq}\approx k_{\rm wn}/(2m)$ the sound speed). 
 % Assuming an IR dominated spectrum as done before, and using
 % the results in ref.~\cite{Gorghetto:2020qws} for the evolution of the spectrum during the non-linear regime after $H=m$, 
Since $k_{\rm wn}\sim {\cal O}(10)k_J$, then $\mathcal{P}=(q_J/k_{\rm wn})^3\sim (k_J/k_{\rm wn})^6={\cal O}(10^{-6})$.
While a dedicated numerical study would be valuable, we suspect that, in this case, isocurvature-induced structures would have negligible observational consequences.
 
 For $N_w>1$, if the network enters the domain wall scaling regime, our estimate
$k_{\rm wn}\sim {\cal A}^{1/3}_w H_d R_{d} \ll m R_d$, is far in the IR with respect to the quantum Jeans scale. It is easy to check indeed that at equality the spectrum is peaked at scales
$\sim q_J$, from which one would conclude that structures of size $2\pi/q_J$ could start forming promptly (see ref.~\cite{Gorghetto:2022ikz} for a recent discussion). However, since these density fluctuations are made
of axions with larger momentum (by a factor of $m/H_d$), free-streaming effects are important.
% \footnote{We thank Mehrdad Mirbabayi for pointing this out.} 
The free-streaming momentum at equality (see e.g.  ref.~\cite{Amin:2022nlh}), in our case turns out to be $k_{\rm wn}/k_{\rm FS}\simeq {\cal A}^{1/3}_w \log(T_d/T_{eq})={\cal A}^{1/3}_w{\cal O}(10)$. The peak is washed out and the largest fluctuations are therefore at a momentum scale $k_{\rm FS}$, with the power spectrum now suppressed by a factor $(k_{FS}/k_{\rm wn})^3={\cal O}(10^{-3})$, at least for small ${\cal A}_w$. The typical size of $\delta \rho/\rho$ at that scale is therefore ${\cal O}(10^{-1}\div10^{-2})$, so that moderately large overfluctuations of the density perturbations could start collapsing not much after equality. 
Precise estimates would require a dedicated numerical study and a better understanding
of the axion spectrum from domain wall decays. What we can say is that those overfluctuations able to collapse at around equality would lead to structures of
size 
\begin{equation}
  R_{\rm halo}\simeq \frac{\pi}{k_{\rm FS}} \simeq
15~{\rm pc}\ \left[\frac{10^{-5}~{\rm eV}}{m}\right]\left[\frac{10^{9}~{\rm GeV}}{f}\right]^2 \left[\frac{g_d}{10}\right]^{\frac13} \frac{\log(T_d/T_{\rm eq})}{10} \,,
\end{equation}
and mass
\begin{equation}
 M_{\rm halo}\simeq \rho_{\rm eq}\frac{4\pi}{3} \left( \frac{2\pi}{k_{\rm FS}}
 \right)^3\simeq
2\cdot 10^8\, M_{\odot} \, \left[\frac{10^{-5}~{\rm eV}}{m}\right]^3\left[\frac{10^{9}~{\rm GeV}}{f}\right]^6 \left[\frac{g_d}{10}\right] \left[\frac{\log(T_d/T_{\rm eq})}{10}\right]^3 \,.
\end{equation}
These estimates vary significantly with $m$ and $f$, even within the parameter space where ALPs explain dark matter. Their impact depends on what fraction goes into these structures, or larger ones later, and whether they survive tidal disruptions. 
It would certainly be interesting to perform more systematic studies,
 as they could further tighten the allowed parameter space and/or provide new opportunities for discoveries.

\subsection{Multi-string scenarios}
\label{sec:multiString}
%%%%%%%%%%%%%%%%%%%%%%%%%%%%%%%
A key feature of the constructions in \cref{sec:AxionsFromWilsonLines,sec:StringTheory} was the existence of \textit{two} symmetry breaking scales $v_\pm$ for a single axion.
In the case $v_\pm \lesssim T_{\rm max}$, the universe undergoes two phase transitions, producing different string species.
The general family of string solutions can be classified by the two independent winding numbers $(n_-,n_+)$ of the $2\pi-$periodic $\theta_-,\theta_+$ \footnote{In section \cref{sec:postInflationaryALPs}, only $(1,0)$ strings were ever present.}. If not that the gauge field lives in the bulk of an extra dimension, the theory is essentially the same as those studied in refs.~\cite{Klaer:2017qhr,Hiramatsu:2019tua,Hiramatsu:2020zlp,Niu:2023khv} and is well approximated by them for $L v_\pm \ll 1$. In short, if $n_\pm \neq 0$, then the corresponding $U(1)_\pm$ is restored at the string core (i.e. $|\Phi_\pm|=0$ is realized there), which gives a contribution $\sim v_\pm^2$ to the string tension. 
Far from the core, the string's field configuration is determined by its axionic global charge $q_a=n_-+n_+$. It looks like a regular axion string with winding $q_a$ and a logarithmically divergent tension $\pi q_a^2v^2\log(v \ell_{\rm IR})$, where typically $\ell_{\rm IR}\sim H^{-1}$ in a scaling regime. We discuss more details of general $(n_-,n_+)$ strings in \cref{app:string_solutions}.

\begin{figure}[t]
        \centering
         \includegraphics[width=0.55\textwidth]{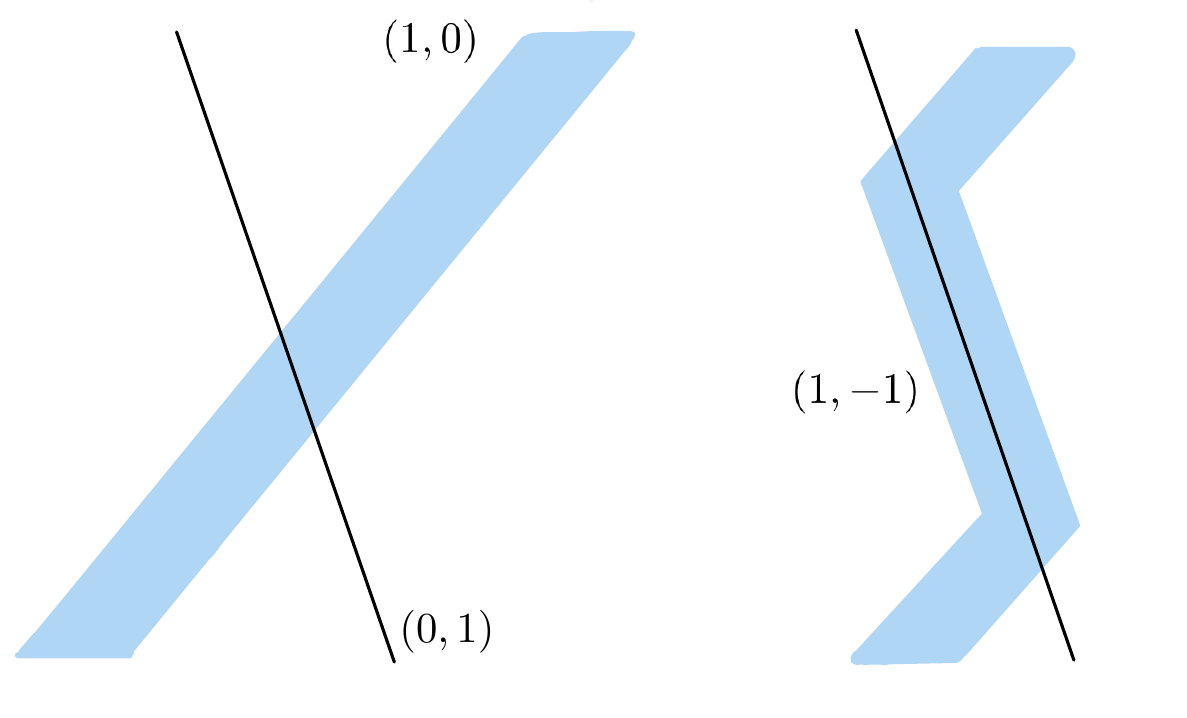}
         %
         %%%%%%%%%%% Caption %%%%%%%%%%
        \caption{The lighter, thicker  $(1,0)$ axion string tends to dress the heavier, thinner $(0,1)$ axion string, leaving the purely local $(1,-1)$ string, a network of which may survive till the present day.
        }
        %%%%%%%%%%%%%%%%%%%%%%%%%%%%%%%
        %
        \label{fig:multistrings}
\end{figure}

Consider first some amount of hierarchy $v\simeq v_-\ll v_+$. At $T\sim v_+$ the first phase transition takes place, the zero mode of $U(1)_5$ is spontaneously broken, but PQ is still preserved while $T\gtrsim v_-$. 
Thus, the $(0,1)$ strings formed look just like familiar local strings in this temperature window, with tension $\mu_{(0,1)} \approx \pi v_+^2$. 
When the second, PQ-breaking, phase transition takes place at $T\lesssim v_-$, the axion becomes a good degree of freedom. From far away, the old $(0,1)$ strings, as well as the newly formed $(1,0)$ ones, both look like regular global-symmetry-breaking axion strings with logarithmically diverging tension $\mu \supset\pi v^2\log\left(v/H\right)$, except that the former still have a much heavier thin core set by $v_+$. On sub-horizon scales,
the two will tend to combine, with the lighter, thicker $(1,0)$ strings `dressing' the $(0,1)$, as depicted in \cref{fig:multistrings}, to form a purely local $(1,-1)$ string\footnote{The opposite combination $(1,1)$ will plausibly be avoided on energetic grounds the same way in the minimal scenario single winding modes do not combine to form double winding modes.}. The existence of the `Y-junction' was already pointed out in \cite{Niu:2023khv}.
For the degenerate case $v_-\approx v_+$ the two axionic strings form at the same epoch, are equally heavy, and rather dress each other. 
Of course strings will also interact and recombine with their own species as usual. 

One obviously important phenomenological question is to what extent does the axion dark matter prediction change compared to that of the minimal scenario of \cref{sec:postInflationaryALPs}. Given the uncertainties already present in the latter, we can only hope to speculate qualitatively for now. 
% While the dressing of \cref{xxx} occurs on sub-horizon scales, at any given time $(0,1)$ and 
It seems quite plausible that the network might evolve to some scaling regime as with single string systems, with roughly comparable numbers per Hubble patch of $(1,0)$, $(0,1)$ and $(1,-1)$ string species. The authors of refs.~\cite{Niu:2023khv,Lu:2023ayc} have suggested a possible enhancement to the final axion abundance in the presence of global strings with a heavy core, since this results in a larger energy emission rate during a scaling regime (by a factor of $\sim v_+^2/f^2$). However, it is also plausible that this enhancement may go only into UV axion modes, and thus not affect the dark matter prediction. It would certainly be interesting to further study this question.

The most robust and phenomenologically novel consequence of this non-minimal post-inflationary scenario appears to be the survival of a  network of local strings with tension set by the heavy scale $v_+$, after the collapse of the global strings. This relic network survives to the present day and may have potentially detectable imprints.
Currently the most robust constraint comes from $G\mu \lesssim 10^{-7}$ from the CMB \cite{Planck:2013mgr,Planck:2015fie} \footnote{Mentioned already in \cref{sec:postInflationaryALPs} in the context of long lived global strings.}, where the tension here is $\mu\approx \pi v_+^2$. The gravitational wave signal from local strings is potentially much stronger, but far less understood due to a much higher sensitivity on the fate and distribution of small loops (as reviewed for example in ref.~\cite{NANOGrav:2023hvm}). 

We note that in all our discussion we have assumed that all masses were set by the appropriate dimensionful scale, with radial mode masses $m_{\rho_\pm}\sim v_\pm$ and $U(1)_5$ (zero mode) vector mass $m_V \approx g_4 v_+ \sim m_{\rho_+}$. This is not the case, for example, in the SUSY II scenario of \cref{sec:Naturalness}.  There, SSB only occurs when an initially SUSY-flat direction is broken by soft terms, resulting in $m_{\rho_\pm} \ll v_\pm \sim m_V$. The dynamics of these deeply Type-I cosmic strings can potentially be far more complex, as studied for example in refs.~\cite{Cui:2007js,Hiramatsu:2013tga}.

%%%%%%%%%%%%%%%%%%%%%%%%%%%%%%%
%%%%%%%%%%%%%%%%%%%%%%%%%%%%%%%

% \paragraph{Supersymmetric comments:} The supersymmetric version of \textit{KK}SVZ was described in \cref{sec:Naturalness}. The single field $\Phi_-$ is swapped for two brane-localized chiral fields, with $\Phi,\Phi'$ their scalar parts, for which the relevant part of the potential is
% \begin{align}
%     V_{\rm I} = \lambda^2 \left|{\Phi} {\Phi}' - {v_-^2 \over 2} \right|^2  \ ,
% \end{align}
% where we drop the `$-$' subscript for simplicity. 
% Below the phase transition $T\lesssim v_-$ the PQ symmetry is spontaneously broken by $ \langle {\Phi} {\Phi}' \rangle = v_-^2/2$. 
% The potential will quickly (anti-)align the respective phases to be $\langle \theta' \rangle = - \langle \theta \rangle$, while a winding in the massless $\theta_-\equiv\theta-\theta'$ variable will result in a topologically stable string configuration. 
% The extra flat direction $|\Phi|=v^2/2|\Phi'|$ does not affect the physics of interest here (moduli do not preclude the existence of topologically stable string solutions \cite{Penin_1996,Achucarro:2001ii,Pickles:2002ym}); ultimately, we imagine it to be lifted by SUSY breaking effects, after which the string network will plausibly look like the minimal scenario.

%%%%%%%%%%%%%%%%%%%%%%%%%%%%%%%%%%%%%%%%%%%%%%%%%%%%%%%%%%%
\section{Conclusions}
\label{sec:Conclusions}
%%%%%%%%%%%%%%%%%%%%%%%%%%%%%%%%%%%%%%%%%%%%%%%%%%%%%%%%%%%

We discussed how extra-dimensional gauge theories with localized charged fields may lead to  
4D theories with high-quality global symmetries\footnote{We focused on \(U(1)\) global symmetries, but the method could be extended to other (non-Abelian) groups.}.  
These manifest themselves as genuine (not accidental) symmetries, meaning that no protection mechanism is present in the 4D EFT to forbid low-dimensional
symmetry-breaking operators, which are instead present but have exponentially suppressed Wilson coefficients.  
The mechanism only requires a small extra dimension (not necessarily much larger than the Planck scale),  
reconciling the lore about the absence of global symmetries in quantum gravity with  
the common practice of imposing them by hand in 4D EFTs.  

In the spontaneously broken phase, these symmetries lead to high-quality NGBs.  
The QCD axion could be one of them. Since the symmetries can be linearly realized  
in 4D, the decay constant scale is unrelated to the compactification scale.  
The symmetry can be completely restored within 4D, allowing  
post-inflationary axion scenarios to be as robust as their pre-inflationary  
counterparts with respect to possible PQ violations in consistent UV completions of gravity. 
We apply the mechanism even to the minimal KSVZ axion model.  
We discussed also how the required conditions are generic in many  
string theory compactifications, which could therefore generate an open-string axiverse, besides the well-known closed-string one.  

The fact that localized fields in extra dimensions could be used to realize field-theoretic axions with high-quality PQ symmetry was already recognized  
in ref.~\cite{Cheng:2001ys}, which constructed a supersymmetric 5D axion model similar to the one presented in section~\ref{sec:5DOrbifold}.  
Compared to \cite{Cheng:2001ys}, we generalized the idea to show how it could be used to protect any linearly realized \(U(1)\) global symmetry in 4D.  
We also demonstrated how the mechanism could generate the minimal KSVZ post-inflationary axion dark matter model,  
quantified the degree of symmetry protection, showed how the same mechanism can be realized 
in string theory compactifications, and justified the underlying assumptions required in 5D.  
As mentioned earlier, QCD axion models from localized open-string states in string theory have been studied in a number of works  
(see, e.g., ref.~\cite{Berenstein:2012eg,Honecker:2013mya,Cicoli:2013cha,Choi:2014uaa}).  
Here, we clarify that the true origin of the emergent global symmetries in 4D lies in the presence of multiple localized charged states,  
rather than in the presence of "anomalous" \(U(1)\) fields and the associated Green-Schwarz mechanism, as is often inferred.  
As a result, the number of effective 4D global symmetries is generally much larger than the number of gauge fields,  
leading to the possibility of a multitude of post-inflationary axions.  
The same mechanism could also lead to non-Abelian effective global symmetries.

Motivating the presence of light axions that underwent post-inflationary evolution highlights new targets and opportunities in the ALP parameter space.
Compared to the pre-inflationary alternative, advantages include greater predictivity and the presence of several purely gravitational probes that complement those from astrophysics and direct searches.
We explicitly examined constraints and signals --- such as those from dark matter and its substructures, dark radiation, isocurvature fluctuations, and gravitational waves --- only for the most minimal and predictive implementation, assuming standard cosmology, a single axion abundance with constant mass, no significant level-crossing effects, etc.
We highlighted qualitatively instead a multi-species cosmic string scenario, which could be generic in the type of constructions presented, leading to distinctive phenomenological features.

The picture in \cref{fig:ALPplot} for the minimal scenario should be considered only a starting point, with significant uncertainties remaining and much room for improvement.
Richer/different phenomenology is also certainly possible. Assumptions of minimality can be relaxed.
The general class of axionic cosmic strings and their cosmological dynamics (as well as other topological defects formed during global-symmetry-breaking phase transitions) should be better understood.
In this post-inflationary world, manifold directions lie open for exploration.

\section*{Acknowledgments}
%%%%%%%%%%%%%%%%%%%%%%%%%%%%%%%%%%%%%%%%%%%%%%%%%%%%%%%%%%%

We thank Mehrdad Mirbabayi for useful discussions. We also wish to acknowledge the California Gym in Roiano, Trieste, where we worked out a non-trivial portion of this paper.

%%%%%%%%%%%%%%%%%%%% APPENDIX %%%%%%%%%%%%%%%%%%%%
\appendix
%%%%%%%%%%%%%%%%%%%%%%%%%%%%%%%%%%%%%%%%%
%\section{Appendix} \label{app:strings}

%%%%%%%%%%%%%%%%%%%%%%%%%%%%%%
\section{More about 5D}
\label{app:5DOrbifold}
%%%%%%%%%%%%%%%%%%%%%%%%%%%%%%

In this appendix we give a few more details regarding the construction
of \cref{sec:5DOrbifold}, in particular for the benefit of the reader less familiar with extra-dimensions. The scalar sector is given explicitly by
% \begin{align}
% \begin{split}
% % \int_{y_+}^{y_-} dy
%         S_{\rm UV} \supset \int d^5x  \, \left[ - {1\over 4}F_{MN}F^{MN} + \{|\mathcal{D}_\mu \Phi_\pm|^2 - V_\pm(|\Phi_\pm|)+y_\pm \Phi_\pm \psi_{a(*)} \psi_{b(*)} + {\rm h.c.} 
%     \} \delta(y-y_\pm) \right] \ , 
%     \label{eq:ToyModel}
% \end{split}
% \end{align}
\begin{align}
% \int_{y_+}^{y_-} dy
        S_{\rm UV} \supset \int d^4x \int^{y_+}_{y_-} dy  \, \left[ - {1\over 4}F_{MN}F^{MN} + \{|\mathcal{D}_\mu \Phi_\pm|^2 - V_\pm(|\Phi_\pm|) 
    \} \delta(y-y_\pm) \right] \ , 
    \label{eq:ToyModel}
\end{align}
where a sum over $\pm$ is implicit, we define the delta functions at the boundaries to integrate to unity, and $\mathcal{D}_\mu = \partial_\mu - i g_5 q_\pm A_\mu(x^M)$, with $q_\pm=\pm 1$ for $\Phi_{\pm}$, 
Choosing $y_\pm=L,0$, a general decomposition of $A_M$, consistent with the stated boundary conditions $A_5(y_\pm)=0$ and $\partial_yA_\mu(y_\pm)=0$, is
\begin{align}
\label{eq:AmuA5decomposition}
% {a^0_\mu\over \sqrt{L}} + 
    A_\mu = \sum_{n=0}^{\infty} {\rm q}^n{a_\mu^n  \over \sqrt{L}} \cos\left(y n\pi \over L\right) \ , \quad  A_5 = \sum_{n=1}^{\infty}{\rm q}^n  {a_5^n  \over \sqrt{L}} \sin\left(y n\pi \over L\right) \ ,
\end{align}
where ${\rm q}^0 = 1$, ${\rm q}^{n>0} = \sqrt{2}$ will ensure canonical normalization.
Plugging these into \cref{eq:ToyModel}, the theory is dimensionally reduced to 4D, with a tower of vectors $a_\mu^{n}(x^\mu)$ and one of NGBs $a_5^{n>0}(x^\mu)$. In the absence of anything else, the latter are all eaten by the KK modes $a_\mu^{n>0}$, while the zero mode $a_\mu^0$ remains massless. We have the more interesting case when symmetry is broken at the boundaries,
\begin{align}
\label{eq:MexicanPotentials}
    V(|\Phi_{\pm}|) = {m_{\rho_+}^2 \over 2 {v_+}^2} 
    \left(|\Phi_\pm|^2 - {v_\pm^2 \over 2}\right)^2 \ , \quad \implies \Phi_\pm = {1 \over \sqrt{2}}(\tilde{\rho}_\pm + v_\pm) e^{i \theta_\pm} \ ,
\end{align}
with $m_{\rho_\pm}$ the masses of the two radial mode excitation.
If only one of $v_\pm$ is non-zero, the corresponding $\theta_\pm$ is also eaten, $a_\mu^0$ becomes massive, and $U(1)_{\rm PQ}$ survives as a global symmetry. When both $v_\pm\neq 0$, we are left with a NGB in 4D --- the axion $a(\theta_\pm,a_5^n)$ defined in \cref{eq:AxionOrbifold}.
This can be made manifest in the dimensionally reduced action, whose quadratic part is
% \begin{align}
% \label{eq:GoldstonesToDiagonalize}
%    -{1\over 4}\sum_{n=0} \left(F^{n}_{\mu\nu}\right)^2 +
%    {1\over 2} \sum_{n=1}^\infty \left( {n\pi \over L} a^n_\mu + \partial_\mu a_5^n \right)^2
%      %
%    +  {v_-^2\over 2} \left(\partial_\mu\theta_- + g_4 \sum_{n=0}^\infty {\rm q}^n a_\mu^n \right)^2 +  {v_+^2\over 2} \left(\partial_\mu\theta_+ -  g_4 \sum_{n=0}^\infty {\rm q}^n a_\mu^n (-1)^n \right)^2  \ .
% \end{align}
\begin{align}
\label{eq:GoldstonesToDiagonalize}
\int d^4x \left[ 
   -{1\over 4}\sum_{n=0} \left(F^{n}_{\mu\nu}\right)^2 +
   {1\over 2} \sum_{n=1}^\infty \left( {n\pi \over L} a^n_\mu + \partial_\mu a_5^n \right)^2
   +  {v_{\pm}^2\over 2} \left(\partial_\mu\theta_{\pm} \mp g_4 \sum_{n=0}^\infty {\rm q}^n a_\mu^n (\mp 1)^n\right)^2 \right] \ ,
\end{align}
by a field transformation $a^n_\mu \rightarrow a^n_\mu + \partial_\mu \lambda^n$, where the $\lambda^n$ are chosen to kill all mixing terms between $a_\mu^n$ and the $a_5^n,\theta_\pm, \lambda^n$. It is not difficult to show that this leads to the conditions
\begin{align}
   -{n \pi \over L}\left( {n\pi \over L} \lambda^n + a_5^n \right) &= 2 \sqrt{2}g_4 v_-^2 \left( \theta_-+ g_4 \sum_{m=0}^\infty {\rm q}^m\lambda^m \right) = 2 \sqrt{2}g_4 v_+^2 \left( \theta_+ - g_4 \sum_{m=0}^\infty {\rm q}^m\lambda^m (-1)^m \right)  \equiv \mathcal{C} \ ,
    \nonumber \\
    n\left( {n\pi \over L} \lambda^n + a_5^n \right) &= 0 \ ,
    % \qquad (n \  \text{even}) \ ,  
    \label{eq:LambdaConditions}
    % \\
    % &= 4g_5 \left( { g_4^2 L^2}
    % + {1\over v_-^2} + {1\over v_+^2} \right)^{-1} \left(\theta_- +  \theta_+ - 2 \sqrt{2} \, g_4\sum_{n \;{\rm odd} }{L \over n\pi}a_5^{n}\right) \nonumber \\
    % & \equiv 4g_5 v^2 (a/v)
    % % \\
    %  % & \equiv C  \ . \hspace{7cm} \ (n \  \text{odd})
    %  \ ,
\end{align}
% \flag{
% \begin{align}
%     \begin{split}
%     \label{eq:lambda_transformations}
%         \lambda^{n } &= - {L \over n \pi} a_5^n \qquad (n>0 \ \text{even}) \ ,\\
%     \lambda^n &= - {L \over n \pi} a_5^n - {2 \sqrt{2} g_4 L^2 \over n^2\pi^2} v a 
%     \qquad (n\ \text{odd})\ ,
%     \\
%         \lambda^0 &=  \sum^\infty_{n>0 \ \rm even} {\sqrt{2}L \over n \pi} a_5^n\ + \frac{1}{2g_4}\left[ \theta_+ - \theta_- + v a \left( {1\over v_-^2} - {1\over v_+^2}\right)\right].
%     \end{split}
% \end{align}
% }
for $n$ odd and even respectively. One can prove by further manipulation of the upper chain of equations in (\ref{eq:LambdaConditions}) that $\mathcal{C}=2\sqrt{2}g_4 v^2(a/v)$, where  $a$ and $v$ were defined in \cref{eq:AxionOrbifold}. Then, by straightforward use of eqs.~(\ref{eq:LambdaConditions}), the quadratic action of \cref{eq:GoldstonesToDiagonalize} reduces to 
\begin{align}
\int d^4x \left[ 
   {1\over 2} \left(\partial_\mu a\right)^2-{1\over 4}\sum^\infty_{n=0} \left(F^{n}_{\mu\nu}\right)^2  +
   {1\over 2} \sum_{n=1}^\infty {n^2\pi^2 \over L^2} \left( a^n_\mu \right)^2
   +  {v_{\pm}^2\over 2} g_4^2 \left(\sum_{n=0}^\infty {\rm q}^n a_\mu^n (\mp 1)^n\right)^2 \right]  \subset S_{4D}\ ,
\end{align}
where the axion, as advertised, is the only NGB left. Again, the sum over $\pm$ is implicit. The tower of massive vectors can now also be diagonalized, though we do not do it here. Suffice it to say that only in the regime $g_4 v_\pm \ll L^{-1}$ is the lightest mass $\simeq g_4^2(v_-^2+v_+^2)$ parametrically below the KK scale.

The explicit forms of the $\lambda^{n}$ can be  easily solved for from eqs.~(\ref{eq:LambdaConditions}) if one so desired. These can be thought of as defining unitary gauge, with a 5D gauge parameter $\Lambda=\sum_{n=0}^\infty \sqrt{{\rm q}^n/L}\, \lambda^n \cos\left( yn\pi/L \right)$. 
In this gauge, $\theta_\pm = (v^2/v_\pm^2)a/v $ and $a_5^n=-(2\sqrt{2}Lg_4v^2/n\pi)a/v$.

The model \cref{eq:ToyModel} was then extended, as per \cref{fig:5DorbifoldSketch}, by the introduction of (all left-handed) colored fermions with Yukawa interactions $\Lag_{\rm UV} \supset \{y_\pm \Phi_\pm \psi_{a(*)} \psi_{b(*)} + {\rm h.c.}\} \delta(y-y_\pm)$ and the necessary Chern-Simons terms in \cref{eq:ChernSimons} to cancel localized gauge anomalies. Under anomalous chiral transformations $\psi_{b(*)}\rightarrow \psi_{b(*)}e^{-i\theta_\pm}$ the $\theta_\pm$ are moved from the Yukawa interactions to Chern-Simons terms and recombine with the KK modes of $A_5$ to give the QCD axion as per \cref{eq:QCDaxion_defining_interaction}. 
% The reader might wonder how the Strong CP problem works in this context. In principle, one can imagine two bare QCD vacuum angles $\theta_{L,R}$, localized on the left (-) and right (+) branes respectively. The combination $\theta_L=-\theta_R$ is gauge equivalent to zero\footnote{It can be cancelled by shifting $A_M$ by the derivative of a constant in the mixed Chern-Simons term in \cref{eq:ToyModel}. \flag{Actually I guess it ends up as theta terms for $U(1)_5$.}}, 
% while  the combination $\theta_L=\theta_R$ is the one dynamically set to zero by the QCD axion.
Apart from this, additional terms in the final axion theory are given by
\begin{align}
 S_{4D} \supset \int d^4x  \,   {v^2\over 2 v_\pm^2}\left(\tilde{\rho}^2_\pm + 2 \tilde{\rho}_\pm v_\pm\right)\left({\partial_\mu a \over v}\right)^2 + {\partial_\mu a \over v} \left({v^2 \over v_-^2}  \bar{\psi}_{b*} \sigma^\mu \psi_{b*} + { v^2 \over v_+^2} \bar{\psi}_b \sigma^\mu \psi_b\right) \ + \text{ KK  }\ ,
\end{align}
where here are using two-component notation for spinors, and
we have left out couplings of the axion to Chern-Simons terms of massive (KK) modes of the bulk gauge fields.
We note that the latter can in general enhance the mass of the axion (without spoiling the Strong CP problem) by the contribution of small (effectively 5D) instantons \cite{Gherghetta:2020keg}, but since the effect is model-dependent, we do not comment on it further here.

% and (covariant)
% kinetic terms for the colored left-handed fermions $\psi_{a,b}$, $\psi_{a*,b*}$ and gluons are implicit. To this we add \cref{eq:ChernSimons} for the cancellation of gauge anomalies.
% (and same for the gluon fields $G^a_\mu$ and $G_5^a$, $a=1,\dots,8$). 

%%%%%%%%%%%%%%%%%%%%%%%%%%%%%%
\section{A Superstring Compactification Example}
\label{app:susystring}
%%%%%%%%%%%%%%%%%%%%%%%%%%%%%%
We give here an explicit string theory example of the $D$-brane configuration
discussed in \cref{sec:StringTheory} consistent with supersymmetry, so that the 
potential vevs of the localized PQ fields could be made parametrically small
with respect to the compactification scale without tuning.

We do not attempt to construct a complete realistic string vacuum solution, since
our purpose here is only to demonstrate that the proposed D~brane configuration
could be made compatible with supersymmetry. Therefore, we will not worry about
global consistency conditions of the configuration (cancellation of global tadpole conditions, Bianchi identities, moduli stabilization, SUSY breaking, SM embedding, etc.). We will assume that other elements in the compactification (such as other $D$-branes, $O$-planes, fluxes) can achieve that in a supersymmetric way (or better with 
some parametrically small SUSY breaking effect, 
which can be parametrized via soft terms).\footnote{In fact, it would
make little sense to try to construct a fully consistent compactification
at this stage without also reproducing the full Standard Model, a positive cosmological
constant, SUSY breaking, baryogenesis, inflation, etc.}
We have however checked that fully consistent SUSY vacua can be found
with the required properties (see below).

For simplicity, we consider a toroidal orientifold compactification of type-IIA string theory with $O$6~planes and intersecting $D$6~branes. Most likely, a realistic compactification would require a much richer manifold. Using the common notation $\prod_{A=1}^3(n_A,m_A)$ to indicate the winding numbers of the 3-cycles wrapped by $D$6~branes and $O$6-planes, we place the orientifold planes parallel to $(1,0)(1,0)(1,0)$.

\begin{figure}[t]
    \centering
    \includegraphics[width=1\linewidth]{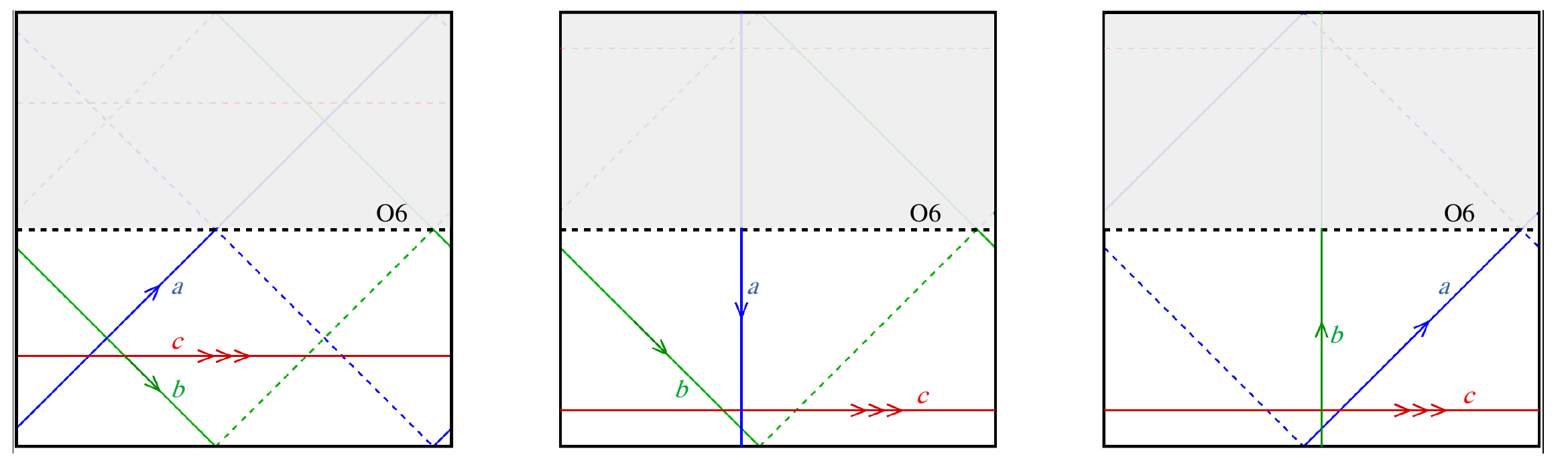}
    \caption{The supersymmetric $D6$-branes and $O6$-planes configuration in the $T^6=T^2\times T^2\times T^2$ compactification reproducing the matter content
    of the model discussed in the text. In particular the blue and green lines corresponds to the two $U(1)_{a,b}$ $D6$-branes, while the red one to the $SU(3)$  color stack of branes. Dashed lines refer to the corresponding $D6$-brane images under the O6 orientifold projection, while the shaded area to the redundant portion of the $T^6$ projected out by the orientifold.}
    \label{fig:branes}
\end{figure}

Let us start considering two branes ($a$ and $b$) at angle as in fig.~\ref{fig:branes}
with the following wrapping numbers
(for a review of this type of constructions see e.g. refs.~\cite{MarchesanoBuznego:2003axu,Uranga:2005wn} 
and references therein):
\begin{equation}
    D6_a:\ (1,1)(0,-1)(1,1)\qquad D6_b:\ (1,-1)(1,-1)(0,1) \ .
\end{equation}
They intersect twice, leading to two chiral superfields ($\Phi_1$ and $\Phi_2$) 
with identical charges [1,-1] w.r.t. the $U(1)_a\times U(1)_b$ gauge fields on the two branes. They are both neutral w.r.t. the diagonal $U(1)_{a+b}$ and have identical charge w.r.t. $U(1)_{a-b}$, 
hence the phase rotations
$\Phi_{1,2} \to e^{\pm i\alpha} \Phi_{1,2}$ correspond to an 
exact global symmetry in 4D (up to exponentially small effects). 

Both $D6_{a,b}$ branes host four additional charged chiral superfields each, localized at 
the intersection with the $O$6 planes and with charges $[2,0]$ and $[0,-2]$ respectively.
These are instead charged under $U(1)_{a+b}$, while sharing again the same charge 
w.r.t. $U(1)_{a-b}$. When considered all together, out of the ten independent $U(1)$ 
phase rotations of the ten localized fields, only two linear combinations are gauged, while all the others
remain as high-quality global symmetries. If the scalar components of these superfields
acquire vevs, they will generate multiple light NGBs. The two $U(1)$ gauge bosons could become massive by either eating two combinations of such NGBs, or
the bulk internal components of the RR 3-form ($C^{(3)}$) gauge field if they have not
been eaten by other $D$6 branes. In any case, a number of NGBs remain uneaten and, being associated to non-local operators on the torus, will remain exponentially light.
We can see already in this very minimal setup how an axiverse could arise from localized charged fields in intersecting brane constructions.

One important property of the construction above is that it is compatible with  supersymmetry. $D$6~branes at angle preserve at least 4 supercharges (corresponding to minimal supersymmetry in 4D) if the sum of the three angles w.r.t. the $O$6~planes vanishes
(mod $2\pi$). In general, this condition is moduli dependent, however, if the compactification preserves supersymmetry, the condition translates into a discrete choice
of winding numbers for the branes. From the 4D effective supergravity point of view,
such a condition corresponds to the cancellation of the $D$ terms coming from field-dependent
Fayet-Iliopoulos (FI) terms (see e.g.~\cite{Villadoro:2006ia} and references therein for
a discussion). In phenomenologically relevant supergravity theories, $D$ terms are always proportional to $F$ terms so that they vanish automatically in supersymmetric compactifications. Note that FI terms are generated as a result of the $D6_{a,b}$ branes gauging part of the shift symmetries of the bulk RR $C^{(3)}$ axion fields, but they cancel
against each other on the vev of the moduli. In the specific example above,
the $D6_{a,b}$ brane configuration is supersymmetric if the moduli are stabilized
supersymmetrically such that $R_1/R_2=R_4/R_3=R_6/R_5$ (where $R_i$ are the sizes of the six main cycles of $T^6$). In supersymmetric flux compactifications, the condition above
is guaranteed if fluxes satisfy the (generalized) Freed-Witten anomaly cancellation
conditions \cite{Freed:1999vc,Villadoro:2006ia} on the $D$6~branes. It is simple to verify that our brane configuration 
can be successfully embedded in a full supersymmetric string compactification where
all moduli are stabilized and all consistency conditions satisfied, such as that found in ref.~\cite{Villadoro:2005cu} and further discussed in ref.~\cite{Camara:2005dc}.

If we now allow for SUSY to be broken at scales parametrically lower than
the compactification scale, the induced soft terms would offset the $D$~term 
cancellation and trigger vevs for the localized charged scalar fields controlled
by SUSY soft terms, hence parametrically small with respect to the compactification scale.

The extension to a KSVZ like 4D theory as the one discussed in \cref{sec:StringTheory} could be
realized by simply adding a stack of 3 $D$6-branes parallel to the $O$6-planes  
(i.e. $D6_c:\ (1,0)(1,0)(1,0)$) playing the role of color branes. These $D6$~branes
intersect each $D6_{a,b}$ and their $O6$ images once leading to 4 chiral supermultiplets
$\psi_a[-1,0,3]$, $\psi_b[0,1,\bar 3 ]$, $\psi_{a^\star}[-1,0,\bar 3]$ and 
$\psi_{b^\star}[0,1,3]$, where the last entry is the representation under $SU(3)$
of the color D6$_c$~brane stack. We have now all the ingredients of the example in \cref{sec:StringTheory} necessary to realize a low energy effective KSVZ model. 
It is easy to find linear combinations of the phase rotations of the localized fields that are not gauged but anomalous under the color group, so that, after the scalar components of the uncolored 
superfields develop a vev, a QCD axion of the KSVZ type arises.  As for the previous
example, also this one can easily be embedded into fully consistent string compactifications such as those in ref.~\cite{Villadoro:2005cu}.

%%%%%%%%%%%%%%%%%%%%%%%%%%%%%%%%%%%%
\section{Cosmic String Solutions}
\label{app:string_solutions}
%%%%%%%%%%%%%%%%%%%%%%%%%%%%%%%%%%%%
 
We describe here in a bit more detail the non-minimal cosmic string solutions that can arise in the models presented in this work, focusing on the case $v_\pm \ll L^{-1}$, as relevant in particular for the discussion in \cref{sec:multiString}. The basic principles of this appendix appear already in the literature (see in particular refs.~\cite{Klaer:2017qhr,Niu:2023khv}).
We start off allowing arbitrary coprime $U(1)_5$ charges $q_\pm$ for $\Phi_\pm$, with $q_\pm=\pm 1$ matching the discussion in the main text\footnote{Extending to non-coprime charges is straightforward, if not somewhat tedious.}.
% We will quote results for general co-prime charges $\mathcal{D}_\mu\Phi_\pm = \left(\partial_\mu - i  e q_\pm A_\mu\right) \Phi_\pm$ while the model highlighted in the main text corresponds to $q_\pm=\pm 1$. 
In the above regime, for our purposes we can focus on the 4D limit of \cref{eq:ToyModel}, keeping only the zero mode gauge field $a^0_\mu$,
% Starting from the model in \cref{sec:5DOrbifold}, and ignoring all KK modes, the dimensionally reduced 4D theory is just
\begin{align}
        \Lag = - {1\over 4} F^0_{\mu\nu}F^{0,\mu\nu} + |\mathcal{D}_\mu \Phi_-|^2 
    + |\mathcal{D}_\mu \Phi_+|^2 
    - V_-(|\Phi_-|)- V_+(|\Phi_+|)
\ ,
\label{eq:TwoAbelianHiggsforString}
\end{align}
where now $\mathcal{D}_\mu = \partial_\mu - i e q_\pm a^0_\mu$, and $e\equiv g_4$ for convenience.
% This will be a good approximation as long as both scales of symmetry breaking $v_\pm$ are smaller than the KK scale. 
In the broken phase (at the level of the approximation of \cref{eq:TwoAbelianHiggsforString}), the vector gains a mass $m^2_V=e^2(q_-^2 v_-^2+q_+^2v_+^2)$, and the axion and its decay constant are given by $a/v = q_+ \theta_- - q_-\theta_+ $ and $v^2 = v_-^2v_+^2  /( q_-^2v_-^2 + q_+^2v_+^2) $.

%%%%%%%%%%%%%%%%%%%%%%%%%%%%%%%%%%%%%%%%%%%%%%
% \flag{As long as $q_+$ and $q_-$ are coprime it might make sense, but otherwise\dots}
%%%%%%%%%%%%%%%%%%%%%%%%%%%%%%%%%%%%%%%%%%%%%%
The string solution ansatz in cylindircal coordinates $(r,\phi,z)$ is
\begin{align}
    \Phi_\pm  = {1\over \sqrt{2}}v_\pm \rho_\pm(r) \ e^{i \theta_\pm(\phi)} \ ,  \qquad   \theta_\pm = n_\pm \phi \ , \qquad
    a^0_\phi =  g(r)/e \ ,
    \label{eq:Usual_ansatz}
\end{align}
where $n_\pm\in \mathbb{Z}$ is the number of windings of the $\theta_{\pm}$ phases. The equations of motion, assuming the Mexican-hat potentials as in \cref{eq:MexicanPotentials}, are
\begin{align}
\label{eq:radial_modes}
    \rho_\pm'' + {1\over r}\rho_\pm' &=  {(q_\pm  g - n_\pm)^2 \over r^2} \rho_\pm  + {m_{\rho_\pm}^2 \over 2} \left(\rho_\pm^2 -1 \right)\rho_\pm \ , 
     \\ 
    g'' -  {g' \over r} &=  2v_-^2 \rho_-^2  e^2 q_- (q_-   g - n_-)+
    2v_+^2 \rho_+^2   e^2 q_+ (q_+   g - n_+) \ .
    \label{eq:vectorEOM}
\end{align}
Regular, asymptotic boundary conditions are given by
\begin{align}
\begin{split}
\label{eq:TwoFieldAsymptotic}
    \rho_{\pm}(r\rightarrow \infty) &\rightarrow 1  \ , \qquad
    g(r\rightarrow \infty) \rightarrow  \frac{n_- q_- v_-^2 + n_+ q_+ v_+^2}{q_-^2 v_-^2 + q_+^2 v_+^2} \ ,
\end{split}
\end{align}
where the value of $g(\infty)$ is such that the right hand side of  \cref{eq:vectorEOM} is zero at large distances.
In this asymptotic regime, the radial mode equations (\ref{eq:radial_modes}) look like a pair of global strings  with effective winding charges (the coefficients of the $r^{-2}$ terms) given by
\begin{align}
        n_{\pm}^{\rm eff} \equiv  q_\pm g(\infty) - n_\pm 
        % = \mp q_a\frac{q_\mp v_\mp^2}{q_-^2 v_-^2 + q_+^2 v_+^2} 
        = \mp q_{\mp} q_a  \frac{v^2}{v_\pm^2}  \ ,
    \label{eq:EffectiveCharges}
\end{align}
where $q_a \equiv (q_- n_+-q_+n_-)$ can be seen as the total axionic charge.
% \begin{align}
%         N^-_{\rm eff} \equiv q_- g(\infty) - n_- =  (q_- n_+-q_+n_-)\frac{q_+v_+^2}{q_-^2 v_-^2 + q_+^2 v_+^2}  \ , \\
%     N^+_{\rm eff} \equiv q_+ g(\infty) - n_+ =
%     -(q_- n_+-q_+n_-)\frac{q_-v_-^2}{q_-^2 v_-^2 + q_+^2 v_+^2} \ . 
%     \label{eq:EffectiveCharges}
% \end{align}
If $q_a \neq 0$, the gradient energies in the scalar field are not screened at large distances and the string tension diverges logarithmically, as for standard global strings \cite{Klaer:2017qhr}. More precisely, from the angular (covariant) gradient $|\mathcal{D}_\phi \Phi_\pm|^2$ terms 
\begin{align}
    \mu  \supset \pi v^2 q_a^2 \int {dr \over r} + \dots \approx \pi v^2 q_a^2 \log\left(m_{\rho_-} \ell_{\rm IR}\right) + \dots 
    \label{eq:TwoFieldLogPiece} 
\end{align}
where $\ell_{\rm IR}$ regulates the IR divergence and  we  assume that $m_{\rho_-}^{-1}\sim v^{-1}$ sets the scale of the outermost region. This contribution to the tension determines the long range interactions between $q_a\neq0$ strings. It is responsible for the dressing discussed in \cref{sec:multiString}, where the $(1,0)$ and $(0,1)$ form bounds states of $(1,-1)$ to minimize the far-field potential energy.

We note, en passant, that the first corrections to the asymptotic limit \cref{eq:TwoFieldAsymptotic} are easily computed
\begin{align}
    \rho_{\pm} \rightarrow 1-{(n_\pm^{\rm eff})^2 \over m_{\rho_\pm}^2 r^2} + \dots \ , \qquad
    g &\rightarrow g(\infty) - 
{q_a^3\over r^2}\frac{2 q_- q_+  v_-^2 v_+^2 \left(m_{\rho_+}^2 q_+^2 v_+^4 - m_{\rho_-}^2 q_-^2 v_-^4 \right)}
{ m_{\rho_-}^2 m_{\rho_+}^2 \left(q_-^2 v_-^2 + q_+^2 v_+^2\right)^4 } + \dots 
    % \frac{n_- q_- v_-^2 + n_+ q_+ v_+^2}{q_-^2 v_-^2 + q_+^2 v_+^2} - {1 \over r^2} 2 q_a^2
    \label{eq:Far_field_solution}
\end{align}
which implies the existence of a magnetic field sourced by the axion winding, which decays as $B_z = {g'(r) / r} \propto 1/r^4 \ ,$ as $r\rightarrow \infty$. To our knowledge, this minor aspect has not been pointed out. We leave the determination of any interesting consequences thereof to future work.

At the opposite extreme, in the inner core of the string near the origin $r=0$, the field profiles consistent with regular boundary conditions go as
\begin{align}
\label{eq:10solutionNearOrigin}
\rho_\pm(r\rightarrow 0) \rightarrow c_{n_\pm} \ r^{n_\pm} \  \ , \qquad
        g(r\rightarrow 0) &\rightarrow c' \  r^2 \ ,
\end{align}
% \begin{align}
% \label{eq:10solutionNearOrigin}
% \rho_\pm(r\rightarrow 0) \rightarrow c_{n_\pm} \ (m_{\rho_{\pm}}r)^{n_\pm} \  \ , \qquad
%         g(r\rightarrow 0) &\rightarrow c' \ (m_V r)^2 \ ,
% \end{align}
where $c_{n_\pm}$ and $c'$ are (dimensionful) constants. The contribution to the string tension from the core in full generality is beyond our purposes. The authors of ref.~\cite{Niu:2023khv} present a general formula derived by variational method. 
What we will say is that, if one of the phases $\theta_\pm$ has zero winding $n_\pm =0$, then $c_{n_\pm =0}=1$ exactly (i.e. the $U(1)_\pm$ symmetry of $\Phi_\pm$ is not restored at the origin when no winding of $\theta_\pm$ is present). For the $(1,0)$ string, the main protagonist of this work, $\rho_+(r\rightarrow0)\rightarrow 1+ \mathcal{O}(r^4)$. 

In the hierarchical regime $v_-\ll v_+$ the $(1,0)$ string should look more and more like a regular global string $\rho_g(r)$. Assuming $\rho_-=\rho_g(r)+\mathcal{O}\left(v_-^2/v_+^2\right)$, while $g(r)$ and $\rho_+-1$ start at most at order $\mathcal{O}\left(v_-^2/v_+^2\right)$, it is not hard to show that $g(r) \sim \rho_g(r)^2 v^2_-/v_+^2 + \mathcal{O}\left(v_-^4/v_+^4\right)$ --- which indeed is consistent with both boundary behaviors \cref{eq:Far_field_solution,eq:10solutionNearOrigin} --- and that the tension $\mu_{(1,0)}$ matches that of a standard global string up to $\mathcal{O}\left(v_-^4/v_+^4\right)$ corrections.

% This results in a contribution to the string tension $\mu \supset \mu_{\rm core(s)} \approx n_\pm^2v_\pm^2$

In summary, for our purposes\footnote{In the discussions in the main text, only $|n_\pm|=0,1$ and $q_\pm =1$ appear.}, the tension of the $(n_-,n_+)$ string is parametrically given by
\begin{align}
    \mu_{(n_-,n_+)} \approx \pi v^2 q_a^2 \log\left(v \ell_{\rm IR}\right) + n_-v_-^2 + n_+v_+^2  \ ,
\end{align}
where we remind that $q_a$ is the axionic charge $q_a = (n_++n_-)$ for the case $q_\pm =\pm 1$.

% \nocite{*}
\bibliographystyle{jhep}
\bibliography{main}
\end{document}